\documentclass[fleqn,usenatbib]{mnras}

\usepackage{newtxtext,newtxmath}

\usepackage[T1]{fontenc}
\usepackage{color}

\usepackage{xcolor}
\definecolor{red}{rgb}{0.96, 0.36, 0.36}

\DeclareRobustCommand{\VAN}[3]{#2}
\let\VANthebibliography\thebibliography
\def\thebibliography{\DeclareRobustCommand{\VAN}[3]{##3}\VANthebibliography}

\usepackage{graphicx}	
\usepackage{amsmath}	
\usepackage{orcidlink}
\usepackage{xcolor}

\title[Stream actions and energies in deforming MW haloes]{Action and energy clustering of stellar streams in deforming Milky Way dark matter haloes}

\author[R. A. N. Brooks et al.]{Richard A. N. Brooks$^{1,2}$\thanks{E-mail: richard.brooks.22@ucl.ac.uk}\orcidlink{0000-0001-5550-2057},
Jason L. Sanders$^{1}$\orcidlink{0000-0003-4593-6788},
Sophia Lilleengen$^{3,4}$\orcidlink{0000-0001-9046-691X}, 
Michael S. Petersen$^{5}$\orcidlink{0000-0003-1517-3935} \newauthor and Andrew Pontzen$^{1}$\orcidlink{0000-0001-9546-3849}
\\
$^{1}$Department of Physics and Astronomy, University College London, London, WC1E 6BT, UK\\
$^{2}$Center for Computational Astrophysics, Flatiron Institute, Simons Foundation, 162 Fifth Avenue, New York, NY 10010, USA\\
$^{3}$Institute for Computational Cosmology, Department of Physics, Durham University, South Road, Durham DH1 3LE, UK\\
$^{4}$Department of Physics, University of Surrey, Guildford GU2 7XH, UK\\
$^{5}$Institute for Astronomy, University of Edinburgh, Royal Observatory, Blackford Hill, Edinburgh EH9 3HJ, UK
}

\date{Accepted XXX. Received YYY; in original form ZZZ}

\pubyear{2024}

\begin{document}
\label{firstpage}
\pagerange{\pageref{firstpage}--\pageref{lastpage}}
\maketitle

\begin{abstract}
We investigate the non-adiabatic effect of time-dependent deformations in the Milky Way (MW) halo potential on stellar streams. Specifically, we consider the MW's response to the infall of the Large Magellanic Cloud (LMC) and how this impacts our ability to recover the spherically averaged MW mass profile from observation using stream actions.
Previously, action clustering methods have only been applied to static or adiabatic MW systems to constrain the properties of the host system.
We use a time-evolving MW--LMC simulation described by basis function expansions. 
We find that for streams with realistic observational uncertainties on shorter orbital periods and without close encounters with the LMC, e.g. GD-1, the radial action distribution is sufficiently clustered to locally recover the spherical MW mass profile across the stream radial range within a $2\sigma$ confidence interval determined using a Fisher information approach. 
For streams with longer orbital periods and close encounters with the LMC, e.g. Orphan-Chenab (OC), the radial action distribution disperses as the MW halo has deformed non-adiabatically. 
Hence, for OC streams generated in potentials that include a MW halo with \textit{any} deformations, action clustering methods will fail to recover the spherical mass profile within a $2\sigma$ uncertainty. 
Finally, we investigate whether the clustering of stream energies can provide similar constraints. Surprisingly, we find for OC-like streams, the recovered spherically averaged mass profiles demonstrate less sensitivity to the time-dependent deformations in the potential. 
\end{abstract}




\section{Introduction}

Within the Local Group, the Milky Way (MW) is undergoing a merger with the Large Magellanic Cloud (LMC)\footnote{See \citet{2023Galax..11...59V} for a comprehensive review detailing the effect of the LMC on the MW.}. The LMC is thought to be on its first pericentric passage\footnote{There are proposed scenarios where the LMC is not on its first passage \citep{2023arXiv230604837V}. Although most features of earlier passages are superseded by the most recent passage at a smaller pericenter.} \citep[][]{2007ApJ...668..949B} and to have a dark matter mass $M_{\mathrm{LMC}}\sim 10^{11}\,\mathrm{M}_{\odot}$. Such a large mass for the LMC is needed to explain many Local Group phenomena: for example, the kinematics of MW satellites \citep{2022MNRAS.511.2610C}; dynamical models of stellar streams \citep{2019MNRAS.487.2685E, 2019MNRAS.485.4726K, 2021ApJ...923..149S, 2021MNRAS.501.2279V}; and the timing argument \citep{2016MNRAS.456L..54P}. The LMC has also been observed to generate significant disequilibrium in the MW gravitational potential:~the displacement of the MW disc, a stellar over-density \citep{2019MNRAS.488L..47B, 2019ApJ...884...51G, 2021Natur.592..534C}, and the reflex motion of the stellar halo \citep{2019MNRAS.487.2685E, 2020MNRAS.494L..11P, Petersen2021, Erkal2021}. The orbit of the LMC is sensitive to the assumed Galactic potential \citep[see fig.~3 of][]{2023Galax..11...59V} and, because the LMC is of considerable mass, it is also subject to dynamical friction \citep{1943ApJ....97..255C}. Current state-of-the-art models of the MW--LMC system account for dynamical friction and the reflex motion of both galaxies \citep[e.g.,][]{2015ApJ...802..128G, 2017MNRAS.464.3825P, 2019MNRAS.487.2685E, 2020ApJ...893..121P, 2020ApJ...898....4C, 2021MNRAS.501.2279V, 2022MNRAS.516.1685D, 2023MNRAS.tmp..536K, 2023MNRAS.518..774L}.

Stellar streams act as kinematic tracers of the underlying dark matter distribution within the Galactic potential. Streams form when satellites, dwarf galaxies or globular clusters, orbiting the MW have their stars tidally stripped. Streams are stringent probes of the MW gravitational potential \citep{1999ApJ...512L.109J, 1999MNRAS.307..495H}; stream members roughly delineate orbits in the host potential \citep{McGlynn1990, Johnston1996, 2013MNRAS.433.1813S}, allowing us to infer the accelerations that the stars experience and hence the host’s gravitational field. The LMC has perturbed streams in the MW, especially those with close encounters \citep[e.g., Orphan-Chenab (OC);][]{2019MNRAS.487.2685E, 2021ApJ...923..149S, 2023MNRAS.518..774L, 2023MNRAS.tmp..536K}. We focus on the OC \citep{2006ApJ...645L..37G, 2006ApJ...642L.137B, 2018ApJ...862..114S, 2019MNRAS.485.4726K} and GD-1 \citep{2006ApJ...643L..17G} stellar streams because they cover a broad radial and angular range of the MW halo with the OC stream having a closer encounter with the LMC. 

The dynamics of streams are most simply described in action-angle coordinates \citep{1999MNRAS.307..877T, 1999MNRAS.307..495H}. Once a star is tidally stripped from the progenitor, its orbital actions are conserved while the angles linearly increase with time in a static or adiabatically invariant potential. Modelling a stream using action-angle variables allows straightforward integration in time \citep{2014ApJ...795...95B}, with the angle variables correlated with the frequencies in potentials close to the true one \citep{2013MNRAS.433.1826S}. Although, time-dependent potentials \citep{2015ApJ...801...98S, Buist2015, Buist2017} and the ‘self-sorting' of streams \citep{Sanders2014, 2014ApJ...795...95B} can complicate these correlations. We omit the angle variables in our modelling and focus on the actions alone as the observable quantity.

The stars in a stream originate from small progenitors and will move along similar orbits, thus the transformation from phase-space positions to action space results in stream members being tightly clustered. When a chosen potential maximally clusters stream members in action space, this is said to reflect the true potential for the system \citep{1999MNRAS.307..495H}. Similarly, the energy clustering of stream members displays the same behaviour. \citet{2012ApJ...760....2P} demonstrated that for separable energy distributions, the associated entropy increases under wrong assumptions about the gravitational potential. 

The first attempts using stellar streams and their action clustering were able to recover parameters of the adopted static potential in which mock streams were evolved \citep{2015ApJ...801...98S, Yang2020}. In turn, the application to observational data using multiple streams was able to set constraints on the enclosed mass of the MW for an assumed static Stäckel gravitational potential \citep{Reino2021}. Multiple streams are often used to nullify any biases on galactic potential parameter fitting due to the orbital phase of streams \citep{Reino2022}. However, any time dependency that is not captured in these static models will subject clustering methods to biases as actions may no longer be conserved \citep{2015ApJ...801...98S}. \citet{2022ApJ...939....2A} provide the most recent effort to accommodate time-evolution in the MW potential. They identify MW-like galaxies in FIRE-2 cosmological simulations and generate populations of stellar streams to maximise action clustering. For the time-evolving MW-like galaxies without any mergers, they find actions remain clustered and stable over dynamical times. However, for a larger merger ($1:8$ mass ratio), there is a temporary decrease in action clustering due to the interaction. Furthermore, highly non-linear perturbations to the potential cause a drift in the radial action distribution \citep{2021MNRAS.508.1404B}.

Due to the merger with the LMC ($\sim1:8$ mass ratio), the potential of the MW has deformed \citep{2023MNRAS.518..774L}. Basis function expansions (BFEs) are used to represent complex systems as linear combinations of simpler functions called basis functions. 
As such, BFEs offer the flexibility to model the deformations captured in $N$-body simulations \citep{Lilley2018a, Lilley2018b, Sanders2020, 2020MNRAS.494L..11P, 2021ApJ...919..109G, 2023MNRAS.518..774L}. The $N$-body simulations of \citet{2023MNRAS.518..774L} infer a BFE description using the \textsc{exp} toolkit \citep{2022MNRAS.510.6201P}. This provides a time-evolving MW system in which stellar streams can be generated. The BFE structure allows exploration of which terms describing the deformation to the MW halo will contribute most to the disruption of clustering of stream members in action spaces. We will consider spherical actions, which are not only useful for perfectly non-deforming spherical systems but also act as a solid base for investigating perturbations to spherical potentials \citep{Pontzen2015}. 

Analysis of the Galactic potential using statistical action clustering methods such as Kullback-Leibler divergence \citep[KLD, or relative entropy;][]{2015ApJ...801...98S}, minimum (Shannon) entropy \citep{2012ApJ...760....2P} or Fisher information \citep[a non-clustering method,][]{2018ApJ...867..101B} are all closely related. The latter represents the Hessian, or curvature, of the relative entropy of a conditional distribution with respect to its parameters. \citet{2018ApJ...867..101B} use the inverse of the Fisher information to determine the Cramér--Rao \citep{Rao1945, Cramer1946} lower bounds on model parameters describing a static MW potential given cold stellar stream observations. To properly constrain the global properties of the Galactic potential, they advocate that many streams should be used simultaneously. However, to capture the complexity of our Galaxy's accretion history with the LMC, a time-dependent model must also be used. We use time-dependent MW and LMC dark matter haloes by employing BFEs to determine the Fisher information on the model parameters (see also, Lilleengen et al.~\textit{in prep}). This extends upon previous Fisher information methods which have assumed static MW potential models for the generation of streams \citep{2018ApJ...867..101B}. We investigate the ability of action clustering methods to recover the MW's spherically averaged mass profile when the temporal evolution could include non-adiabatic behaviour. The flexibility of BFEs allows us to easily investigate behaviour for a wide range of deforming MW--LMC potentials.

The plan of the paper is as follows. Sec.~\ref{sec:methods} describes our methodology containing: an overview of BFEs and spherical action-angle coordinates and the framework to generate stellar streams. In Sec.~\ref{sec:action-angles-TD} we present the action distributions of mock stellar streams in various deforming potentials. In Sec.~\ref{sec:results}, we outline our information theory approach and determine the ability to constrain the spherically-averaged MW mass profile in Sec.~\ref{sec:results}. We discuss the results plus any caveats in Sec.~\ref{sec:discussion} and summarise our findings in Sec.~\ref{sec:summary}.

\section{Methods}\label{sec:methods}

In Sec.~\ref{sec:bfes}, we summarise the approach taken in \citet{2023MNRAS.518..774L} to generate their $N$-body model of the MW dark matter halo using the basis function expansion software suite, \textsc{exp} \citep{2022MNRAS.510.6201P}. Plus we outline the expected result when the potential has non-adiabatic behaviour. In Sec.~\ref{sec:sphaction}, we outline the action-angle variables for stream members in spherical potentials. This includes details of the use of the high-performance numerical computing python package \textsc{jax} \citep[][]{jax2018github} in analysing streams, Sec.~\ref{sec:jaxauto}. Finally, in Sec.~\ref{sec:dynmodels}, we present the dynamical modelling used to generate streams.

\subsection{Basis Function Expansions}\label{sec:bfes} 

\subsubsection{\textsc{exp}}\label{sec:exp} 

To generate stellar stream models in a time-evolving MW--LMC system, we need a description of the potential and forces at any arbitrary position and time. Static potentials fail to capture deformations to the MW and LMC dark matter haloes. BFEs offer a framework to describe these deformations. They track the density, gravitational potential and forces as the system evolves over time. BFEs have previously been seen to accurately describe flexible models of the MW \citep{2016MNRAS.463.1952P, 2018ApJ...858...73D, 2019MNRAS.490.3616P, 2020MNRAS.494L..11P, 2021ApJ...919..109G}. In this work, we use the BFEs of the MW--LMC system presented in \citet{2023MNRAS.518..774L} that are simulated using \textsc{exp} \citep{2022MNRAS.510.6201P}, with the expansion coefficients recorded at each time step. All potentials we consider exclude contributions from the MW disc.

The BFE technique uses appropriately chosen biorthogonal density-potential pairs of basis functions, $\{ \varrho_{\mu}(\textbf{x}), \phi_{\mu}(\textbf{x}) \}$, that solve Poisson's equation i.e., $\nabla^{2}\phi_{\mu}(\textbf{x}) = 4\pi G\varrho_{\mu}(\textbf{x})$ and satisfy the biorthogonality condition $\int \mathrm{d}^3\mathbf{x}\,\phi_{\mu}(\textbf{x})\varrho_{\nu}(\textbf{x}) = 4\pi G\delta_{\mu \nu}$ where $\delta_{\mu \nu}$ is the Kronecker delta. Each basis function, labelled by the index $\mu$, adds a degree of freedom to the system and has an associated coefficient $A_{\mu}$, which determines its contribution to the total description of the system, i.e. the summation over all basis function terms. A system at any given time is described by the basis functions and the coefficients that weight them. Mathematically, the density, $\rho$, and gravitational potential, $\Phi$, are:

\begin{equation}\label{equ1}
    \rho(\textbf{x}, t) = \sum_{\mu}\,A_{\mu}(t) \varrho_{\mu}(\textbf{x}),
\end{equation}

\begin{equation}\label{equ2}
    \Phi(\textbf{x}, t) = \sum_{\mu}\,A_{\mu}(t) \phi_{\mu}(\textbf{x}),
\end{equation}
\noindent
where the basis coefficients are time-dependent and the basis function keeps its fixed functional form.

Basis functions are selected to reflect the system they describe. To model density profiles, $\rho(r, \phi, \theta)$, with deviations away from spherical symmetry, the spherical harmonics $Y_{l}^{m}$ are chosen to describe the distribution in the angular coordinates $(\phi, \theta)$, whilst \textsc{exp} describes the radial dependence (index $n$) by the eigenfunctions of a Sturm-Liouville equation \citep{Weinberg1999}. Each spherical basis function is then represented by the triplet of indices $\mu \equiv (n, l, m)$. The radial index, $n$, determines the number of nodes in the radial basis function. For $l=0$, $n$ equals the number of nodes in the radial function. For $l>0$, there are $n+1$ radial nodes. A maximum truncation in the expansion for the radial part, $n_{\mathrm{max}}$, and angular part, $l_{\mathrm{max}}$, corresponds to a spherical coefficient set of size $(l_{\mathrm{max}}+1)^{2} \cdot (n_{\mathrm{max}}+1)$. The \textsc{exp} method creates a lowest-order monopole term, $\rho_{000}(r)$, that exactly matches the unperturbed input potential-density pair. All other, higher-order, terms are perturbations around the input distribution. If the lowest-order monopole does not match the input pair, more terms are needed to approximate the input distribution.
Another example of a BFE is the classical Hernquist-Ostriker basis set \citep{1992ApJ...386..375H} which expands upon the Hernquist density distribution \citep{1990ApJ...356..359H} as $\rho_{000}(r)$. Alternative choices of analytic basis functions have been made such that the underlying density distribution allows axisymmetric, triaxial, and lopsided distortions \citep{Lilley2018a, Lilley2018b}. 

\subsubsection{\emph{N}-body models and Basis Function Expansions}\label{sec:NbodyMWH} 

An efficient lightweight python interface, \textsc{mwlmc}, has been developed to facilitate the \textsc{exp} simulations of the \citet{2023MNRAS.518..774L} MW--LMC system. This user-friendly interface is publicly available at:~\url{https://github.com/sophialilleengen/mwlmc}. This MW--LMC system is constructed with three components with separate BFEs: the MW dark matter halo, the MW stellar disc and the LMC dark matter halo. The \textsc{exp} method explicitly uses the BFE for the force evaluations in the $N$-body evolution. We describe the MW and LMC dark matter haloes in this section. Descriptions of the BFE and $N$-body models for the MW disc can be found in secs.~2.1 \& 2.2 of \citet[][]{2023MNRAS.518..774L}, respectively. The $N$-body models of \citet{2023MNRAS.518..774L} self-consistently include the effect of dynamical friction on the LMC as it falls into the MW's potential.
Throughout this work, we analyse the deformations of the MW halo. The LMC is described by its full basis expansion throughout.

The LMC dark matter halo is modelled by a Hernquist \citep{Hernquist1990} profile with $M_{\mathrm{LMC}} = 1.25\times10^{11}\,\mathrm{M}_{\odot}$, $r_{s} = 14.9\,\mathrm{kpc}$. This halo is realised with $10^7$ particles and simulated using \textsc{exp} \citep{2022MNRAS.510.6201P} with $l_{\mathrm{max}} = 6, n_{\mathrm{max}} = 23$ \citep{2023MNRAS.518..774L}. 
 
The MW dark matter halo profile is selected from table A1 of \citet{2019MNRAS.487.2685E} as the best fit spherical potential, label ‘sph. rMW+LMC'. A Navarro-Frenk-White profile \citep[NFW;][]{1996ApJ...462..563N} is used to describe the MW halo with $M_{\mathrm{vir}} = 7.92\times10^{11}\,\mathrm{M}_{\odot}$, $r_{s} = 12.8\,\mathrm{kpc}$ and $c=15.3$. This profile is truncated as $\rho_{\mathrm{halo}}(r) = 0.5\rho_{\mathrm{NFW}}(r)(1 - \mathrm{erf}[(r - r_{\mathrm{trunc}})/w_{\mathrm{trunc}}])$ where $r_{\mathrm{trunc}} = 430\,\mathrm{kpc}$ and $w_{\mathrm{trunc}} = 54\,\mathrm{kpc}$. 
This halo is realised with $10^7$ particles and simulated using \textsc{exp} \citep{2022MNRAS.510.6201P} with $l_{\mathrm{max}} = 6, n_{\mathrm{max}} = 17$ \citep{2023MNRAS.518..774L}. 
For the MW halo, it is convenient to describe individual harmonic subsets of $l$. The $l=0$ terms are called the monopole, $l=1$ is the dipole, $l=2$ is the quadrupole, etc. The live simulation of the MW--LMC system begins at $t=t_{\mathrm{live}} = -2.5\,\mathrm{Gyr}$, with present-day at $t = 0\,\mathrm{Gyr}$. At the start of the live simulation, the MW and LMC haloes are totally distinct, with the LMC outside the virial radius of the MW at a distance of $450\,\mathrm{kpc}$.
The density, force, and potential fields before the start of the live simulation have the basis coefficients set to their initial values prescribed at $t_{\mathrm{live}}$. 

\subsubsection{Evolution in increasingly complex systems}\label{sec:inc-complex-potentials}

\begin{figure*}
    \centering
    \includegraphics[width=\linewidth]{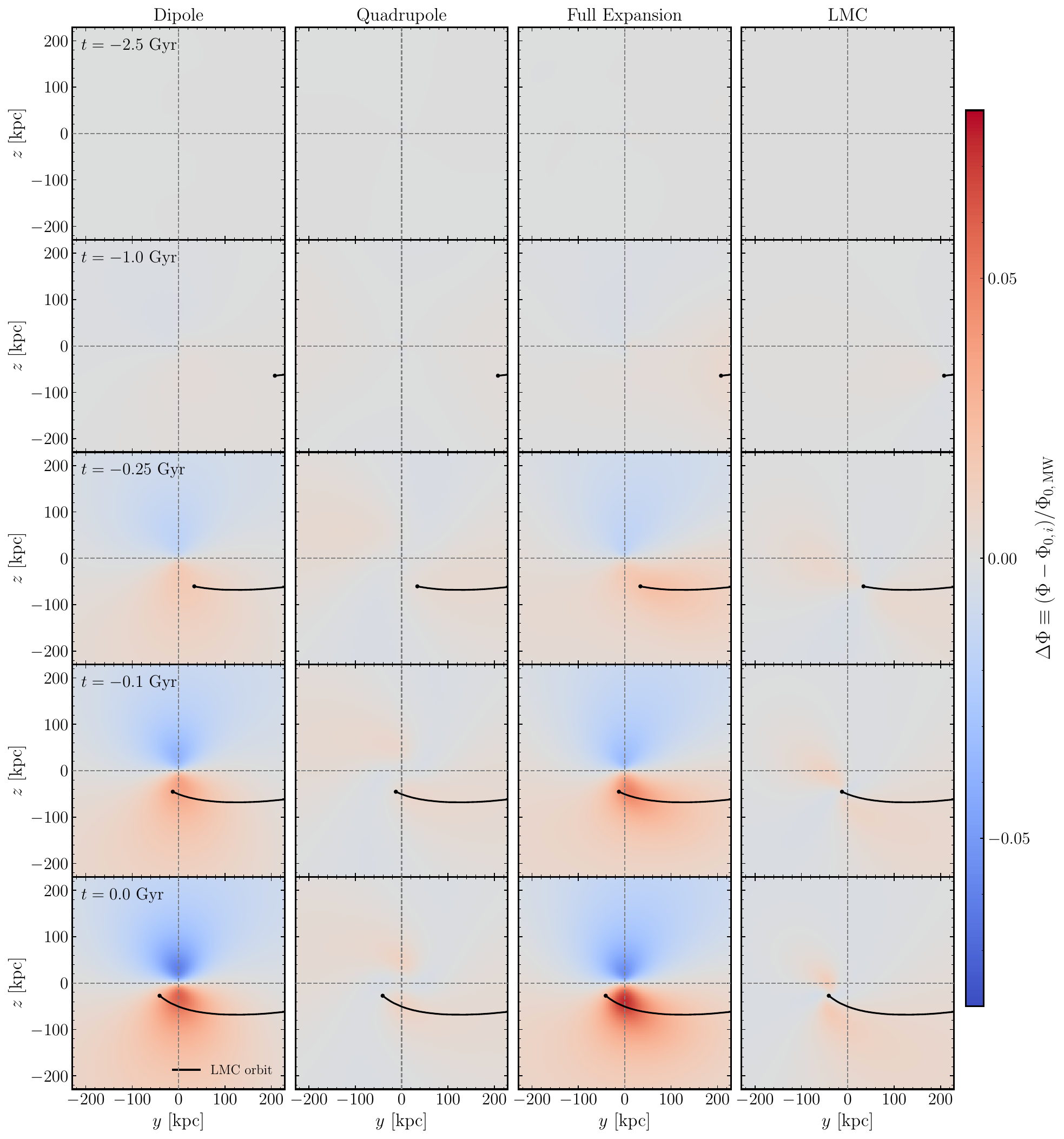}
    \caption{Temporal development of various MW dark matter halo harmonics and the LMC dark matter halo over the live simulation time; $t = -2.5 \,\mathrm{Gyr}$ to $t = 0 \,\mathrm{Gyr}$ with time increasing top to bottom. Going left to right across the columns shows the MW dipole, MW quadrupole, MW full expansion harmonic, and LMC full expansion. The potentials are computed in the $x = 0$ Galactocentric plane in a slab of 10 kpc thickness. The colour map represents the potential contrast, $\Delta\Phi \equiv (\Phi - \Phi_{0,\,i}) / \Phi_{0,\,\mathrm{MW}}$, where $\Phi_{0, i}$ corresponds to the monopole potential computed using only the $l=0$ order of either the MW (first three columns) or LMC expansion (final column). The track of the LMC through this plane is shown as the black line. Halo deformations due to the MW disc are omitted as they are subdominant with respect to the outer halo deformations. A video version of this figure is available at \url{https://www.youtube.com/watch?v=i18zbNxNyf8}. A similar version of this figure using the dark matter densities can be found in appendix~\ref{app:A}.}
    \label{fig:harmonics-potential}
\end{figure*}

To investigate how different harmonic subsets of the full BFE affect the generation of a stellar stream, harmonic terms can be selectively turned off, i.e. by setting all relevant BFE coefficients to zero, to isolate the contributions to the total BFE description of the system. This effect on the OC stream track relative to that of the full BFE expansion is visualised in fig.~5 in \citet{2023MNRAS.518..774L}. They isolate the effect of each term in the BFE by keeping either the MW or LMC live and varying the harmonic contributions of the other. For the MW halo, the largest effect on the OC stream track is from including the dipole harmonic. In this work, we consider 6 MW--LMC potentials to generate streams that all use the full LMC halo BFE but with different harmonic subsets of the MW halo BFE: ‘static monopole' which uses the unperturbed set of monopole coefficients\footnote{This differs from \citet{2023MNRAS.518..774L} who used the set of monopole coefficients at $t = 0\,\mathrm{Gyr}$ for their ‘static monopole' potential.} i.e. before the live simulation starts, ‘evolving monopole', ‘monopole + dipole', ‘monopole + quadrupole', ‘monopole + dipole + quadrupole' and ‘full expansion'. We use the same LMC description for all generated streams so that we can focus on the systematic effect of deformations to the MW halo.

The harmonic orders of the BFE will develop over the entire simulation. At the beginning of the live simulation, $t = t_\mathrm{{live}}$, and for all prior times, there has yet to be any response of the MW's dark matter halo due to the passage of the LMC. At these times, the MW halo can be fully described by its monopole harmonic subset as we do not include the MW disc which would create some halo deformations. Nevertheless, these deformations would be subdominant with respect to the outer halo deformations. 
The in-fall of the LMC, as the satellite galaxy, onto the MW, as the central galaxy, generates density wakes \citep{1943ApJ....97..255C}. The classical `conic' wake trailing the LMC is described as the \textit{transient response}, whereas the response elsewhere in the MW halo is the \textit{collective response} \citep{2021ApJ...919..109G}. These effects will also be reflected in the gravitational potential as the density and potential are related by Poisson's equation.
In Fig.~\ref{fig:harmonics-potential} we demonstrate the temporal development of the MW halo potential contrast for both isolated harmonic subsets and the full basis expansion simulation in the MW--LMC simulations of \citet{2023MNRAS.518..774L}. We show only the harmonic subsets that are considered in this paper, i.e. harmonic orders above the octupole, $l = 3 $, are not shown. Additionally, we show the potential contrast for the LMC halo described by the full basis expansion in the right-most column.
The potential contrast is defined as $\Delta\Phi \equiv (\Phi - \Phi_{0,\,i}) / \Phi_{0,\,\mathrm{MW}}$, where $\Phi_{0,\,i}$ corresponds to the monopole potential computed using only the $n=0$ order of either the MW (first three columns) or LMC expansion (final column). As the system evolves towards the present day (going top to bottom of Fig.~\ref{fig:harmonics-potential}), the amplitude of the potential contrast of all harmonic subsets increases. The MW dipole potential contrast that is generated is stronger than that of the MW quadrupole. This is expected as the MW dipole deformation is known to have the largest effect on the OC stream track \citep{2023MNRAS.518..774L}. The potential contrast of the full expansion LMC is more localised than that of the MW, while also being weaker by a factor of $\sim 4-5$, consistent with the expectation of in-falling satellites \citep{Weinberg1989}. Analysis using the dark matter densities, with comparison to the similar yet distinct MW--LMC BFE of \citet{2021ApJ...919..109G}, can be found in appendix~\ref{app:A}.

\subsection{Actions in spherical potentials}\label{sec:sphaction} 

An integral of motion, $I_{i}(\textbf{x}, \textbf{v})$, is any function of phase space coordinates, $(\textbf{x}, \textbf{v})$, that is a constant along an orbit. The \textit{action-angle} variables, $(\mathbf{J}, \boldsymbol{\Theta})$, use a particular set of canonical coordinates with the three momenta are integrals called \textit{actions} and the conjugate coordinates are the \textit{angles}. This choice of coordinate system makes the Hamiltonian independent of the angle variables, i.e.,  $H = H(\textbf{J})$, so the angles increase linearly in time. As the actions are conserved quantities on bound orbits, the full orbit can be explored by varying the angles only, $\boldsymbol{\Theta}$; the orbital three-tori \citep{arnold1989mathematical, 2008gady.book.....B}. The actions quantify the rotation around the symmetry axis, the oscillation amplitude in the radial direction, and the direction perpendicular to the symmetry axis.

Following \citet{2008gady.book.....B} we define the radial action:

\begin{equation}\label{equ3}
    J_{r} = \frac{1}{\pi}\int_{r_{p}}^{r_{a}}\mathrm{d}r\sqrt{2\left(E - \Phi(r)\right) - \frac{L^{2}}{r^{2}}},
\end{equation}

\noindent 
where $L$ is the angular momentum, $E$ is the energy, $\Phi$ is the gravitational potential and integral limits are the orbital perihelion, $r_p$ and aphelion, $r_a$. The other two actions are the azimuthal action $J_{\phi} = L_{z}$ and the latitudinal action $J_{\theta} = L - |L_{z}|$. This completes the triplet of actions, $\mathbf{J} = (J_r, J_{\phi}, J_{\theta})$. When variations in the potential are slow compared to the typical orbital frequencies, $\Omega$, these potentials are labelled adiabatic \citep[][sec.~3.6]{2008gady.book.....B}. The actions of particles in an adiabatic potential are constant and for this reason, the actions are called adiabatic invariants. 

Throughout this work, we calculate actions in spherical potentials only. Often, there will be asymmetry in our chosen MW--LMC potential to generate a stream. We discuss the process of spherically averaging the potential in Sec.~\ref{sec:MLE-estiamtion-monopole-coeffs}. There are very few instances where analytic solutions for Equ.~\eqref{equ3} exist, therefore requiring us to make a numerical estimate. To do this, we have implemented a numerical version of Equ.~\eqref{equ3} in \textsc{jax} \citep{jax2018github} by approximating the integral as a Gauss--Legendre summation over radial bins in the interval between the peri-- and apocentres. To check whether our numerical implementation is successful, we perform a check against radial actions calculated for a mock OC stream generated over 3 Gyr in an analytic isochrone MW potential \citep{Henon1959a, Henon1959b}, with a total mass $M_{\mathrm{MW}} = 10^{12}\,\mathrm{M}_{\odot}$ and scale radius $r_s = 15\,\mathrm{kpc}$, using \textsc{agama} \citep[][]{2019MNRAS.482.1525V}. We find an agreement between the analytic and numerical action calculation of $\sim10^{-6}$~per~cent. 

\subsubsection{\textsc{jax} automatic differentiation}\label{sec:jaxauto}

A key part of our formalism to analyse streams requires the knowledge of the phase-- and action--space derivatives with respect to the quantities which parametrize them. These derivatives are useful in the context of information theory, Sec.~\ref{sec:fisherinfo}, and maximum likelihood estimation, Sec.~\ref{sec:MLE-estiamtion-monopole-coeffs}.
We choose \textsc{jax} to implement these derivatives because it employs automatic differentiation. The premise of automatic differentiation exploits the fact that for any given algorithm, it will execute elementary arithmetic operations, e.g. addition, multiplication, division, and functions e.g. sine, cosine, log. By repeatedly applying the chain rule to these operations, the partial derivatives up to an arbitrary derivative order can be calculated automatically. 

To be able to calculate the derivatives of e.g. the potential and forces from the BFE code, we wrap functions from the \textsc{mwlmc} package in a \textsc{jax} environment. Once wrapped, we can automatically differentiate functions with respect to their input parameters. Important derivatives to obtain are those of the potential. The derivative with respect to the position is simply the negative of the force at that position, $\partial \Phi(\textbf{x})/ \partial \textbf{x} = -\boldsymbol{F}(\textbf{x})$. From Equ.~\ref{equ2}, the derivative of the potential with respect to a time-varying basis function coefficient is:

\begin{equation}\label{equ4}
    \frac{\partial \Phi(\textbf{x}, t)}{\partial A_{\mu}}\Bigg|_{t}  = \phi_{\mu}(\textbf{x}),
\end{equation}

\noindent
which is simply the basis function corresponding to the coefficient evaluated at a given position. An advantage of the automatic differentiation framework is that any subsequent function that depends on the function with calculable derivatives will also have its derivatives automatically calculated. For example, the radial action $J_{r}(\mathbf{x}, \mathbf{v}, \{A_{\mu}\})$ has automatic derivatives with respect to the basis function coefficients, $\partial J_r/ \partial A_{\mu}$, because it is a function of the potential where we know the coefficient derivatives.

To determine the accuracy of the automatic differentiation scheme, we perform a check against derivatives calculated numerically from finite differencing for $10^4$ test particles in the MW halo described by the full BFE. We find the numerical and automatic derivatives for the potential with respect to the coefficients differ by no more than $\sim10^{-9}$~per~cent, while the radial action shows $\sim10^{-1}$~per~cent differences. Taken in comparison to the stream dispersions of $\sim20$~per~cent, this accuracy is more than sufficient for our problem. By using the \textsc{jax} \texttt{grad} function to calculate the derivatives automatically, we are able to find the derivatives with respect to the positions, velocities and coefficients for both the potential and radial action across all stream particles. For the $10^4$ test particles, calculating the derivatives for the radial action over 3 position elements, 3 velocity elements and 882 coefficients takes $\lesssim30\,\mathrm{s}$ when executing the code in a local environment.

\subsection{Dynamical modelling of stellar streams}\label{sec:dynmodels} 

\begin{figure}
    \centering
    \includegraphics[width=\linewidth]{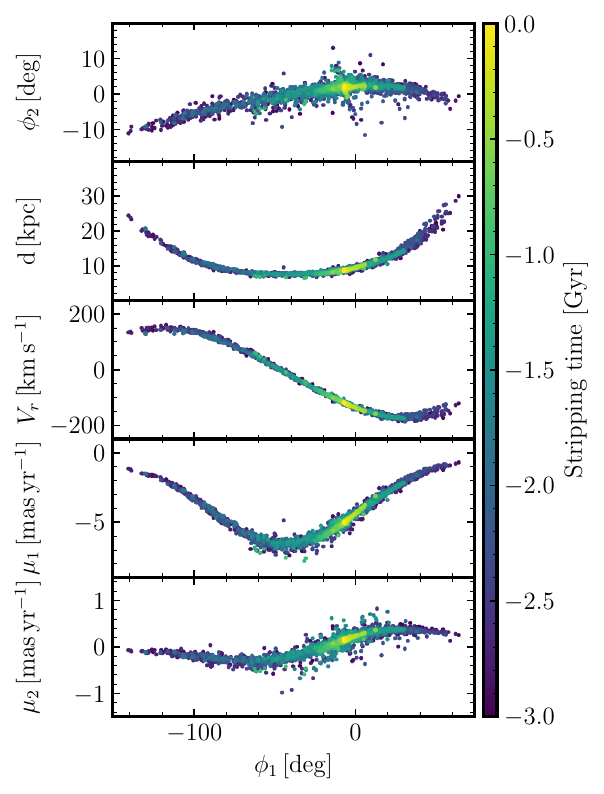}
    \caption{A mock GD-1 stream generated in an MW + LMC potential described by the full basis sets for each dark matter halo. The rows show the stream sky coordinates \citep{2010ApJ...712..260K}, heliocentric distance, radial velocity, and reflex-corrected proper motions, respectively. The colour scale shows the time at which the stream particle was released during the forward integration of the progenitor relative to the present day.}
    \label{fig:stream_obsuncert}
\end{figure}

\subsubsection{Stream generation}\label{sec:mockstream-generation}

To produce realistic models of stellar streams, we use a ‘modified Lagrange Cloud Stripping' (mLCS) technique \citep{Kupper2012, Bonaca2014, 2014MNRAS.445.3788G, 2015MNRAS.449.1391B, 2015MNRAS.452..301F}. Modifications were developed to include the forces from the LMC and reflex motion of the MW in \citet{2019MNRAS.487.2685E}. The stream progenitors are modelled as Plummer spheres \citep{1911MNRAS..71..460P} with initial masses and scale radii as defined in Sec.~\ref{sec:mwstreams}. From the progenitor's present-day position, we rewind the phase-space orbit for $3\,\mathrm{Gyr}$ in a chosen MW--LMC potentials; see Sec.~\ref{sec:inc-complex-potentials}. The system is subsequently forward-evolved in the same potential, and stream particles are released from the progenitor's Lagrange points, $r_{\mathrm{prog}} \pm r_t$, at each time step. The Lagrange, or tidal, radius is found by,

\begin{equation}
    r_{t} = \left( \frac{\mathrm{G}M_{\mathrm{prog}}(t)}{\omega^{2} - \frac{\mathrm{d}^2\Phi}{\mathrm{d}r^2}}   \right)^{1/3},
\end{equation}

\noindent
where $\omega$ is the angular velocity of the progenitor with respect to the MW and $\frac{\mathrm{d}^2\Phi}{\mathrm{d}r^2}$ is the second derivative of the MW potential along the radial direction. We model the mass loss of progenitors as linearly decreasing in time since the progenitors are not seen in observational data \citep{2023MNRAS.tmp..536K}. We account for the velocity dispersion of the progenitor, $\sigma_{\mathrm{v}}$, by randomly drawing velocities from a 3D isotropic Gaussian centred on the velocities of the stripped particles, $\textbf{v}_{\mathrm{strip}}$, with standard deviation $\sigma_{\mathrm{v}} = \sqrt{\mathrm{G}M_{\mathrm{prog}}(t)/(r_t^2 + a_s^2)^{1/2}}$, where $a_s$ is the scale radius of the progenitor. The radial component of $\textbf{v}_{\mathrm{strip}}$ is the same as the progenitor, while the tangential components are set to those at the point halfway between the progenitor and the Lagrange point. We use the same right-handed coordinate system as \citet{2023MNRAS.518..774L} with the Sun’s position at $\textbf{x} = (-8.249, 0,
0)\,\mathrm{kpc}$ and its velocity $\textbf{v} = (11.1, 245, 7.3)\,\mathrm{km}\,\mathrm{s}^{-1}$. We include the self-gravity of the progenitors during the mLCS such that stream particles experience forces due to the progenitor. 

At each time step during the forward evolution of the system, we compute the forces acting on each particle. In the same fashion as \citet{2019MNRAS.487.2685E} and \citet{2023MNRAS.518..774L}, motivated by the results of \citet{Dehnen2011}, we implement an adaptive time step such that computational efficiency and precision are achieved during the integration. To capture the orbit around the MW, we calculate $\Delta\,t_{i,\mathrm{MW}} = \eta \sqrt{\frac{r_{i}}{|\mathbf{a}_{i}|}}$ where $i$ is the index over stream members, $r_i$ is the distance of each particle to the MW centre, $\mathbf{a}_{i}$ is the acceleration each particle feels due to the combined MW and LMC haloes, and $\eta = 0.01$. To capture the orbit around the progenitor, we compute $\Delta\,t_{i,\mathrm{prog}} = \eta \sqrt{\frac{r_{i, \mathrm{prog}}}{|\mathbf{a}_{i, \mathrm{prog}}|}}$ where $r_{i, \mathrm{prog}}$ is the distance of each particle to the progenitor and $\mathbf{a}_{i, \mathrm{prog}}$ is the acceleration each particle feels due to the progenitor. We then determine the minimum time step over all particles $\Delta\,t = \mathrm{min}_{i} (\Delta\,t_{i,\mathrm{MW}},\,\Delta\,t_{i,\mathrm{prog}})$ with a minimum allowed time-step of $0.5\,\mathrm{Myr}$. Similar to \citet{2023MNRAS.518..774L}, including an LMC time-step criterion makes no observable difference to the stream. 

We make the connection to observations of stellar streams as follows. Having generated and evolved a stream through the total integration time, we take a random sample of the stream particles to match the number of likely stream members. The OC stream includes 360 likely members \citep{2023MNRAS.tmp..536K} based on the combination of the southern stellar stream spectroscopy survey \citep[$S^5$,][]{2019MNRAS.490.3508L}, Apache Point Observatory Galactic Evolution Experiment 
 \citep[APOGEE,][]{2017AJ....154...94M}, \textit{SDSS} and Large Sky Area Multi-Object Fiber Spectroscopic Telescope \citep[\textit{LAMOST},][]{2012RAA....12.1197C} survey data. To be conservative, we assume that only 250 members are observationally confirmed and generate a random sample of the idealised OC streams to match this number. Meanwhile, the GD-1 stream has 1155 likely members \citep{Viswanathan2023} identified using \textit{Gaia} Early Data Release 3 \citep{GaiaCollaboration2021, Babusiaux2023}, of which 783 are main sequence stars and we choose to generate random samples to match this value as a conservative estimate for the number of observationally confirmed stream members. 

\begin{figure*}
    \centering
    \includegraphics[width=\linewidth]{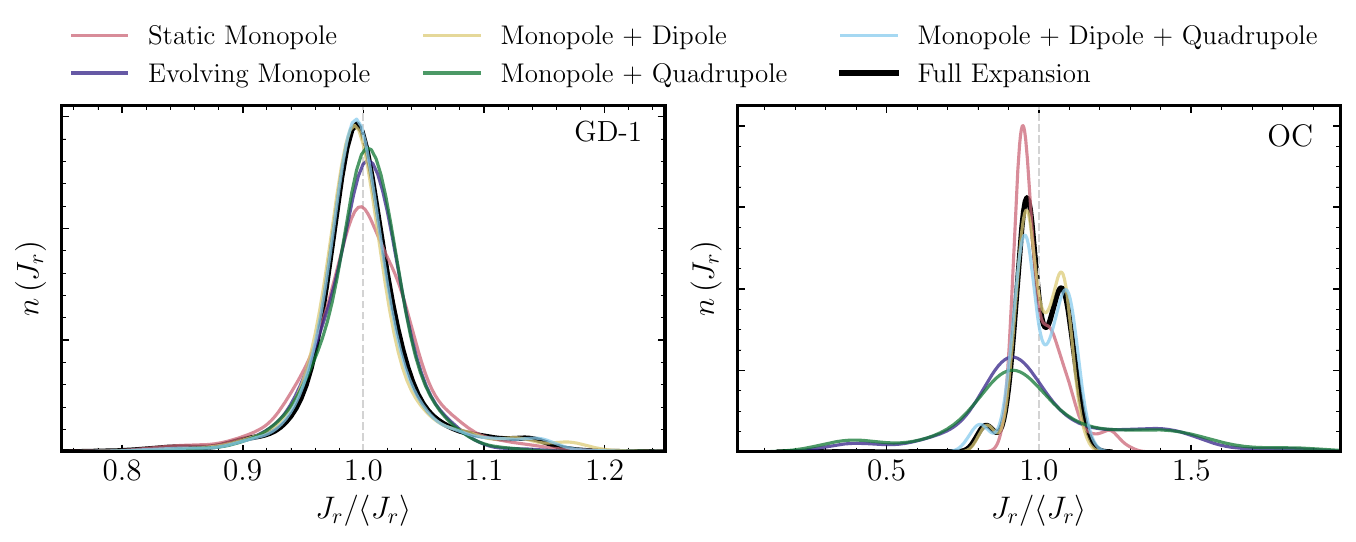}
    \caption{The distribution of radial actions $n(J_r)$ for the GD-1 (left) and OC (right) streams generated in the full basis expansion LMC halo potential plus the MW halo described by six distinct BFE subsets: static monopole (pink), evolving monopole (indigo), monopole + dipole (yellow), monopole + quadrupole (green), monopole + dipole + quadrupole (cyan) and the full expansion (thick black). The spherical radial action is calculated using the spherically averaged MW halo potential. 
    For each mock stream, the distribution is normalised by the mean radial action and shown using a kernel density estimate.
    A radial action distribution that appears non-Gaussian suggests non-adiabatic behaviour in the potential with respect to the stream generated. For GD-1, all action distributions appear Gaussian implying adiabatic potential behaviour locally. For OC streams generated in deforming potentials, they show non-Gaussian action distributions, implying non-adiabatic behaviour locally.}
    \label{fig:Jr_dists}
\end{figure*}

\subsubsection{Milky Way Streams Selection}\label{sec:mwstreams}

 We choose two MW streams with distinct radial ranges and proximities to the LMC such that we infer information about the underlying density/potential fields across different parts of the MW. These streams are:

\begin{enumerate}
    \item Orphan-Chenab (OC): The Orphan and Chenab streams were discovered separately \citep{2006ApJ...645L..37G, 2006ApJ...642L.137B, 2018ApJ...862..114S} before being discovered that they formed two parts of the same stream \citep{2019MNRAS.485.4726K}. This confusion was due to the Chenab part of the OC stream being actively perturbed by the LMC \citep{2019MNRAS.487.2685E}. The OC stream is very long, extending radially $\sim 15 - 80\,\mathrm{kpc}$ with sections passing close to the LMC, making it ideal to investigate the MW and LMC potentials \citep{2023MNRAS.tmp..536K}. To match observational constraints, we model it as a Plummer sphere \citep{1911MNRAS..71..460P} with an initial mass of $M_{\mathrm{prog}} = 10^{7}\,\mathrm{M}_{\odot}$, and a scale radius of $1\,\mathrm{kpc}$ \citep{2019MNRAS.485.4726K}. 
    We set the progenitor's present-day location using the same initial conditions as \citet{2023MNRAS.518..774L}: $\phi_1 = 6.340^{\circ}$, $\phi_2 = - 0.456^{\circ}$, $d = 18.975\,\mathrm{kpc}$, $v_r = 93.786\,\mathrm{km}\,\mathrm{s}^{-1}$, $\mu_{\alpha} = -3.590\,\mathrm{mas}\,\mathrm{yr}^{-1}$, and $\mu_{\delta} = 2.666\, \mathrm{mas}\,\mathrm{yr}^{-1}$, following the notion of \citet{2019MNRAS.485.4726K} and \citet{2019MNRAS.487.2685E}.
    The stream track coordinates $(\phi_1, \phi_2)$ are given in a coordinate system provided by \citet{2019MNRAS.485.4726K}. Appendix B of \citet{2019MNRAS.485.4726K} gives the rotation matrix for this coordinate transformation.
    \item GD-1: The GD-1 stream was discovered in the Sloan Digital Sky Survey \citep[\textit{SDSS},][]{2000AJ....120.1579Y} as a very thin and long, $\sim 63^{\circ}$ structure \citep{2006ApJ...643L..17G}. The progenitor for GD-1 is unknown and has likely fully dispersed. We model the progenitor as a Plummer sphere with an initial mass of $M_{\mathrm{prog}} = 2\times10^{4}\,\mathrm{M}_{\odot}$, and a scale radius of $5\,\mathrm{pc}$. The total stellar mass of the observed GD-1 stream is
    estimated to be $1.8\times10^{4}\,\mathrm{M}_{\odot}$ \citep{deBoer2020}, hence our choice
    of $M_{\mathrm{prog}}$ is above the lower bound for the initial mass of the GD-1 progenitor. 
    We load the present-day 6D phase space position of the GD-1 progenitor in \citet{2019MNRAS.485.5929W} via the \textsc{galpy} module \citep{2015ApJS..216...29B}. The progenitor's initial
    conditions are: $\phi_1 = -39.640^{\circ}$, 
    $\phi_2 = -0.493^{\circ}$, 
    $d = 7.485\,\mathrm{kpc}$, 
    $v_r = 6.337\,\mathrm{km}\,\mathrm{s}^{-1}$, 
    $\mu_{\alpha} = -13.097\,\mathrm{mas}\,\mathrm{yr}^{-1}$, 
    and $\mu_{\delta} = -3.248\, \mathrm{mas}\,\mathrm{yr}^{-1}$. The stream track coordinates $(\phi_1, \phi_2)$ are given in a coordinate system provided by \citet{2010ApJ...712..260K}.
\end{enumerate}

Fig.~\ref{fig:stream_obsuncert} shows a mock GD-1 stream generated in a MW--LMC potential described by the full basis set for each halo. The colour gradient represents the time at which stream particles were released from the Lagrange points relative to the present-day, $t = 0\,\mathrm{Gyr}$. 
 
\begin{figure*}
    \centering
    \includegraphics[width=\linewidth]{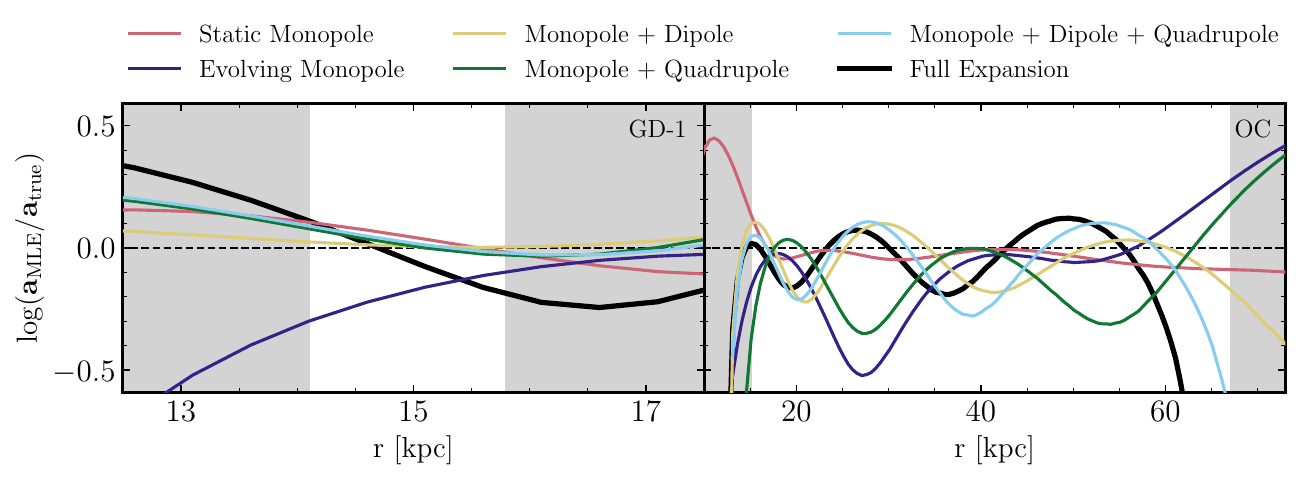}
    \caption{Logarithmic ratio of the spherically averaged BFE acceleration profile, described by the maximum likelihood estimates of the MW halo monopole coefficients, to the original spherically averaged MW halo acceleration profile. \textit{Left:} For a GD-1 stream generated in a full basis expansion LMC plus a MW halo potential described by the following harmonic subsets: static monopole (pink), evolving monopole (indigo), monopole + dipole (yellow), monopole + quadrupole (green), monopole + dipole + quadrupole (cyan) and the full expansion (thick black). \textit{Right:} The same as the left panel but for an OC stream. The grey-shaded region indicates regions outside the average $10^{\mathrm{th}} - 90^{\mathrm{th}}$ percentiles of the radial distribution of stream particles.}
    \label{fig:MLE}
\end{figure*}

\section{Radial actions distributions of streams in time-evolving potentials}\label{sec:action-angles-TD}

We adopt the following process to demonstrate how the radial action distribution of stream members changes in the different time-evolving MW--LMC potentials. We take the positions and velocities of all stream particles at the present-day snapshot. The spherical radial action is calculated using Equ.~\ref{equ3} by using the spherically-averaged potential, i.e. described by the monopole subset of the MW halo BFE only. The resulting 1D distribution of radial actions for the GD-1 (left) and OC (right) streams are shown in Fig.~\ref{fig:Jr_dists}. To compare the distributions for streams generated in different MW--LMC potentials, we have normalised each distribution by its mean radial action.

The actions of a stream specify its path through phase space. The stream members share similar orbits to their progenitor, hence sharing similar actions \citep{1999MNRAS.307..495H, Helmi2020, Deason2024}. As action variables are adiabatic invariants \citep{2008gady.book.....B},  any changes to the potential that are slower than the typical orbital frequency of a stream will retain a Gaussian distributed, or at least well-clustered, set of actions over time \citep{Eyre2011, Sanders2016}. This is because stream stars, before a perturbation to the Galaxy e.g. by the merger with the LMC, will initially share similar orbits (but at different phases) and will still share similar orbits after any slow changes to the potential are complete.

Streams will interact with the infalling LMC, dynamically altering them. These interactions are adiabatic if the present-day distribution of stream actions is clustered. For GD-1-like streams, left panel of Fig.~\ref{fig:Jr_dists}, no matter the degree of time-dependence allowed in the MW halo, the action distribution remains clustered at present-day. Conversely, OC-like streams, right panel of Fig.~\ref{fig:Jr_dists}, are significantly affected by the inclusion of \textit{any} time dependence in the MW halo potential. Spherical actions computed for streams generated in potentials that allow an evolving MW halo monopole or quadrupole show long tails to their distributions. Moreover, the MW halo dipole manifests as a bi-modality in the action distribution. The non-Gaussian nature of these action distributions suggests that the MW halo has deformed non-adiabatically with respect to the OC stream.
As we are evaluating the spherical action, there may be contributions to the action evolution due to non-spherical changes in the potential. We explore the distinction between non-adiabatic and non-spherical contributions to the action evolution in Appendix.~\ref{app:B}. Briefly, for the OC stream generated in the full expansion MW--LMC potential, we find the action evolution is dominated by non-adiabatic changes to the potential. Whereas for GD-1, non-adiabatic contributions are insignificant compared to contributions from the non-spherical changes to the potential.
For both streams generated using a static monopole MW halo potential, the distributions are well clustered as expected. Indeed, the spreads of these distributions are dominated by the intrinsic properties of tidal stripping as opposed to the perturbative effects of the LMC, i.e. $\sigma_{J,\,\mathrm{exc.\,LMC}} / \sigma_{J,\,\mathrm{inc.\,LMC}} \sim 1$.

We note that GD-1 has been associated with perturbations due to the Sagittarius dwarf galaxy merger \citep{Bonaca2020, 2022MNRAS.516.1685D}. The simulations considered in this work do not include Sagittarius and future work will determine whether the GD-1 -- Sagittarius interaction causes its action distribution to disperse further.

\subsection{Actions as adiabatic invariants}

The theoretical attraction of using action variables is their property of adiabatic invariance. This means, that for a time-dependent system, the energies of particles will not be conserved. 
However, for a slowly varying system, there exists a combination of energy and time-dependent parameters, which make up the actions, that remain approximately constant \citep{Vandervoort1961, Landau1969, Wells2007, 2008gady.book.....B}. For our Galaxy, when the potential is assumed as static or slowing-evolving, the actions of a stream remain approximately constant and clustered. 
In this instance, action clustering methods are appropriate to infer global system properties e.g. mass profiles. For an MW--LMC system described by BFEs, the conditions under which actions remain clustered are outlined in appendix~\ref{app:B}. We consider changes in radial action as a function of lookback time for two neighbouring particles evolved in a time-dependent MW--LMC potential compared to the static potential. 
Our Equ.~\ref{equ:B6} demonstrates this, connecting changes in the potential to changes in actions. This equation highlights how each basis function coefficient will affect the action evolution uniquely.
Namely, the ratio $\dot{A}_{\mu}(t) / \Omega$ is a global indicator for adiabatic invariance in the actions. The bracketed terms modulate this global quantity to the location of the stream particles, with the first term measuring the change in the actions of the particles orbiting in a time-dependent system, while the second term measures the change in actions around a particle's orbit in a static system.
In general, non-adiabatic potentials translate themselves to a total change in actions, $\Delta\,J_{r}(t) \sim \mathcal{O}(\sigma_{J_r})$, i.e. comparable to the spread of the original distribution of actions.

\section{Results}\label{sec:results}

\subsection{Information theory}\label{sec:fisherinfo} 

The Fisher information \citep{Fisher1925} is a way of measuring the amount of information that a random variable $\mathbf{y}$ carries about an unknown parameter $\mathbf{x}$ of a distribution that models that random variable. \citet{2018ApJ...867..101B} developed an information framework for cold stellar streams in static potentials where the random variables $\mathbf{y}$ are the tracks of the stream observations (on-sky track, distance, radial velocity and proper motions) while the model $\mathbf{x}$ includes parameters for the progenitor, the baryonic potential components and the dark matter potential components. 

For a more general case where there are $N$ model parameters $\mathbf{x} = [x_1, x_2, ..., x_N]^{\mathrm{T}}$ that describe the variable $\mathbf{y}$, the Fisher information is given by the $N\times N$ positive semi-definite matrix called the \textit{Fisher Information Matrix} (FIM) with the elements: 

\begin{equation}\label{equ7}
   [I(\mathbf{x})]_{i,j} = \mathbb{E}\left[ \left( \frac{\partial}{\partial x_i} \log f(\mathbf{y};\mathbf{x}) \right) \left(\frac{\partial}{\partial x_j} \log f(\mathbf{y};\mathbf{x}) \right) \bigg| \,\mathbf{x}\right],
\end{equation}

\noindent
where $[I(\mathbf{x})]_{i,j}$ is the information that the variable $\mathbf{y}$ carries about the covariances between the model parameters $x_i$ and $x_j$. The probability distribution for the random variable $\mathbf{y}$ conditioned on the value of the model parameters, $\mathbf{x}$, is labelled $f(\mathbf{y};\mathbf{x})$. In our case, we are calculating the present-day information that a radial action distribution of a stellar stream carries about the set of BFE coefficients that are used to model the Galactic potential. Inverting the FIM returns the matrix of Cramér--Rao lower bounds \citep{Rao1945, Cramer1946}. The square roots of the diagonal elements are the bounds on the individual coefficient model parameters. The Cramér--Rao lower bounds are interpreted as the lower bounds for the best-case uncertainties given the data and their uncertainties.

The radial action $J_r$ is calculated starting from the phase-space coordinates of stream members and the model parameters as the BFE coefficients $A_{\mu}$ which define the potential, i.e. $J_r = J_{r}(\mathbf{x}, \mathbf{v}, \{A_{\mu}\})$. For the radial action, we assume a Gaussian distribution centred on the mean action $\langle J_{r}\rangle$ with a standard deviation, $\sigma_{J_r}$, i.e. $f(J_r;A_{\mu}) = \mathcal{N}(\langle J_{r}\rangle,\,\sigma_{J_r}^{2})$. For adiabatic invariant systems, this is a good approximation \citep{Eyre2011, Sanders2016}. 

To account for the stream observational uncertainties, we take the uncertainties for the GD-1 stream based on values given in \citet{GaiaCollaboration2018}, \citet{Malhan2019}, \& \citet{2022MNRAS.516.1685D}, while for the OC stream, we use the values given in \citet{2023MNRAS.tmp..536K}. Each stream member's positions and velocities are convolved with these uncertainties. Using these observation-like positions and velocities, we calculate the actions of each stream, $J_r$, and the spread of the action distribution, $\sigma_{J_r}^{2}$.

Assuming the actions are separable, the BFE-relevant elements of the FIM for the $(a^{\mathrm{th}}, b^{\mathrm{th}})$ coefficients are given by:

\begin{equation}\label{equ8}
    \left[I\left(\mathbf{A_{\mu}}\right)\right]_{a,b} = \sum_{\mathrm{stars}} \frac{\left(\phi_{a} - \langle\phi_{a}\rangle\right) \left(\phi_{b} - \langle\phi_{b}\rangle\right) \left(J_{r} - \langle J_{r}\rangle\right)^{2}}{\Omega_{r}^{2}\,\sigma_{J_r}^{4}}
\end{equation}

\noindent
where $\phi_{j}$ are the basis functions evaluated at the particles' positions, and with the frequency $\Omega_r$. This expression looks similar to minimum entropy methods developed in \citet{2012ApJ...760....2P} as will be discussed in Sec.~\ref{sec:discussion}. A full derivation and discussion of the radial action FIM expression for the general and Gaussian distribution cases can be found in Appendix \ref{app:C}.

\subsection{Spherically averaged Milky Way acceleration profiles}\label{sec:MLE-estiamtion-monopole-coeffs}

Our derivation of the information using the radial action assumes the potential to be spherical. Hence, we only use the Cramér--Rao lower bounds of the spherically averaged monopole BFE coefficients as these terms are independent of angular contributions. For all generated streams, we seek the combination of monopole coefficients that best describe the MW halo potential given the action distribution of the stream generated in the various time-dependent potentials. We employ a maximum likelihood estimation (MLE) to achieve this by minimizing a log-likelihood function for the radial action that is assumed to be Gaussian \citep{Eyre2011, Sanders2016}, i.e:
\begin{equation}\label{equ:8}
    \ln{\left(f(J_r)\right)} = -\frac{1}{2}\left(\ln{\left(2\pi\sigma_{J_r}^2\right)} + \frac{\left(J_{r} - \langle J_{r}\rangle\right)^{2}}{\sigma_{J_r}^2}\right)
\end{equation}

\noindent
The returned spherically averaged coefficients are then used to re-evaluate the potential, the actions plus their derivatives and the Fisher information for the streams generated in time-evolving systems. 

In Fig.~\ref{fig:MLE} we show the logarithmic ratio of the acceleration field described by the MLE coefficients to the ‘true' spherically averaged MW acceleration field described by the original monopole coefficients. In the left (right) panel, we show the ratio across the stream radial range for the GD-1 (OC) streams. The grey-shaded region indicates regions outside the average $10^{\mathrm{th}} - 90^{\mathrm{th}}$ percentiles of the radial distribution of stream particles. In both cases, the acceleration field is only well recovered across the stream range \citep{2018ApJ...867..101B}. 
For GD-1, the expected acceleration profile across the majority of the stream can be recovered reasonably well for all potentials in which we generate streams.
For OC, only in the static monopole case can the acceleration be recovered well across the radial stream range. All other cases that include the dipole and/or quadrupole harmonic of the MW halo BFE description demonstrate large deviations from the expected acceleration profile across the stream range implying we cannot recover useful information about the MW when it has undergone non-adiabatic evolution.

\begin{figure*}
    \centering
    \includegraphics[width=1\linewidth]{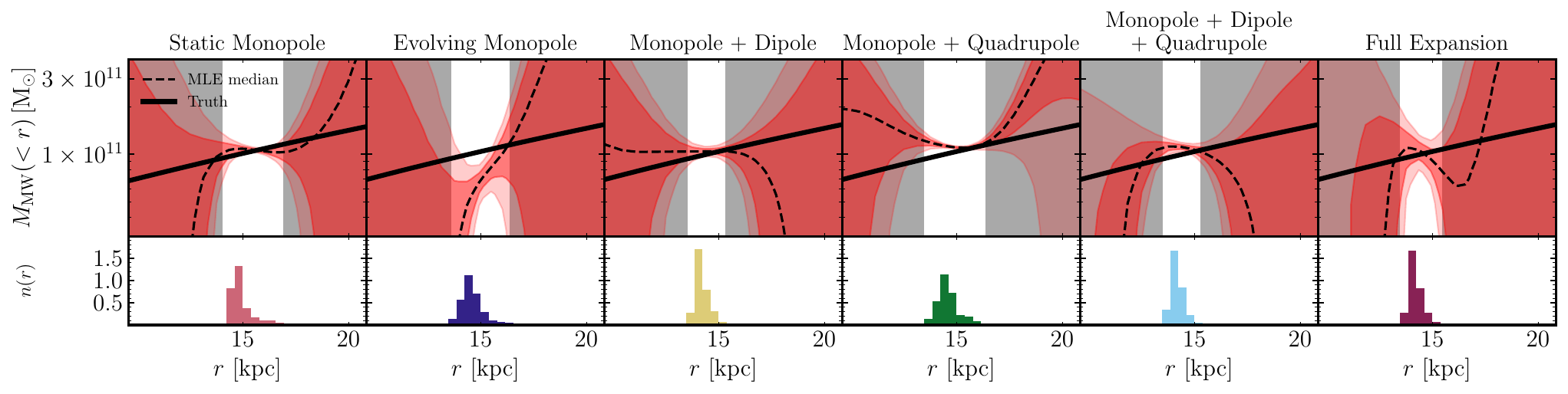}
    \includegraphics[width=1\linewidth]{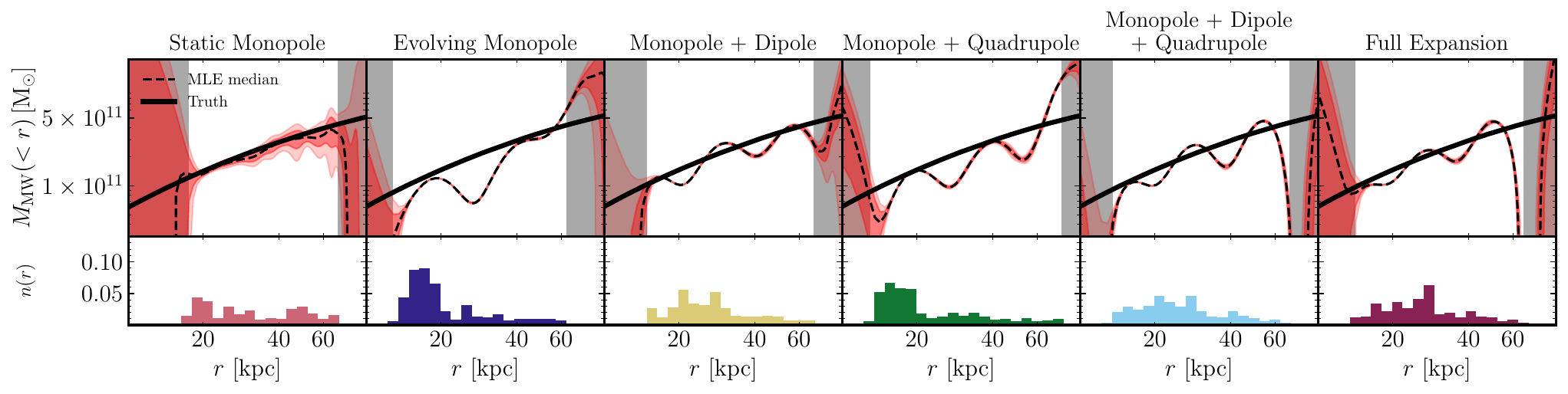}
    \caption{Constraints on the spherically-averaged Milky Way mass profile using the generated GD-1 streams (top row) and generated OC streams (bottom row) from their \textbf{radial action distribution}. The resulting median mass profile (black dashed lines) is shown with the $1\,\sigma$ (dark red) and $2\,\sigma$ (light red) confidence intervals as shaded bands. The number density of stream members as a function of radius is shown in the lower panels, the total number of stream members matches observational counts in Sec.~\ref{sec:mockstream-generation}. The shaded-grey regions indicate radial regions where there are no stream members. The thick black line represents the true MW halo monopole potential governing the spherical part of the basis function expansion. Left to right shows these mass profiles for the streams generated in the fully live LMC + MW halo harmonic subset potentials: static monopole, evolving monopole, monopole + dipole, monopole + quadrupole, monopole + dipole + quadrupole and the full expansion. Streams that cannot recover the mass profile across their radial extent imply their actions have been impacted by non-adiabatic behaviour in the underlying potential.}
    \label{fig:Mass-enclosed-profiles}
\end{figure*}

\subsection{Action clustering - Milky Way mass profile}\label{sec:obslike-constraints}

We now investigate recovering the MW mass profile using action clustering. Using the MLE spherically averaged monopole coefficients determined in Sec.~\ref{sec:MLE-estiamtion-monopole-coeffs}, we calculate the radial actions and quantities required to determine the FIM; Sec.~\ref{sec:fisherinfo}, Equ.~\ref{equ8}. To make the FIM realistic in connection with observation, we make conservative matches for the counts of likely stream members, Sec.~\ref{sec:mockstream-generation}. To avoid biases from outlier stream particles, we select our random samples from the distribution of particles within the $10^{\mathrm{th}} - 90^{\mathrm{th}}$ distance percentile. Once the FIM is known, we take its inverse to return the Cramér--Rao matrix. We draw random samples of spherically averaged coefficients from a multivariate normal distribution with the mean being the MLE spherically averaged coefficients and the covariance matrix being the Cramér--Rao matrix. Using these samples of spherically averaged coefficients in combination with the force basis function weights we compute the radial acceleration profile. We can then infer the spherically averaged mass profiles for the GD-1 streams (top row Fig.~\ref{fig:Mass-enclosed-profiles}) and OC streams (bottom row Fig.~\ref{fig:Mass-enclosed-profiles}). For an assumed spherically symmetric potential, the acceleration and mass are related, $a(r) = \partial\Phi/\partial r = G M(<r) / r^{2} $. In Fig.~\ref{fig:Mass-enclosed-profiles}, the median mass profiles are shown as the dashed black line with $1\,\sigma$ (dark red) and $2\,\sigma$ (light red) confidence intervals. If the system in which the stream was generated is adiabatic, we expect to be able to recover the true MW mass profile (thick black line) across the radial range where there are stream members. The lower panels in both rows of Fig.~\ref{fig:Mass-enclosed-profiles} show the number density, normalised by bin width, of stream members as a function of radius. 

For the mock GD-1 streams, top row Fig.~\ref{fig:Mass-enclosed-profiles}, we can recover the mass profiles within $2\,\sigma$ across the stream range in all cases. Although the stream has visited smaller and larger radii on its orbit, the action clustering method is only sensitive to local accelerations. Hence, we are only locally constraining the flexible BFE description of the mass profile across the radial extent of the stream. Outside this range, there is a dearth of information and the confidence intervals widen. This is in contrast to static parameterizations of potential models which can lead to constraints being placed on regions outside of the stream range \citep[e.g.][]{2019MNRAS.487.2685E, Malhan2019, 2023MNRAS.tmp..536K}. These results imply that any time dependence in the MW--LMC potential is adiabatic over the evolution of a GD-1-like stream. Hence, for GD-1-like streams, we can use the clustering of actions to infer the MW mass profile.

For the mock OC streams, bottom row Fig.~\ref{fig:Mass-enclosed-profiles}, we can recover the MW mass enclosed profile, within $1\,\sigma$ across the stream range, in the static monopole case. This result is expected as the MW halo is not deforming, i.e. it is time-independent, and so the potential will be adiabatic during the evolution of the OC stream.
However, the inclusion of \textit{any} time-dependence in the MW halo potential leads to an inability to recover the mass profile across the stream range. This suggests that the deformations of the MW halo introduce non-adiabatic behaviour in the potential that is sustained throughout the evolution of OC-like streams. This effect is reflected as the nonphysical negative mass dips seen in Fig.~\ref{fig:Mass-enclosed-profiles}. These perturbations to the mass profile mean we are not able to recover the expected profile within the confidence intervals. 
Given the definition of BFEs, if the coefficients assign ‘extra' weight to specific fine-tuning higher radial orders with smaller periodicity, the combination with the basis function weights can generate negative masses. These negative masses should be addressed in future work, although this is a non-trivial exercise. This could be achieved by putting constraints on regions of the coefficient parameter space which permit negative masses to exist. Indeed, allowing negative masses has likely improved the returned mass profile constraints. Nevertheless, these results imply that for OC-like streams created in MW haloes which are time-dependent due to the merger with the LMC, we are unable to use action clustering methods to recover the mass profile as the potential is non-adiabatic and action clustering is no longer preserved. 
We note, for mock OC streams generated in potentials with an evolving monopole or monopole + quadrupole MW halo, the mass profile is biased low. A possible source of this bias is seen in these streams' radial action distributions (Fig.~\ref{fig:Jr_dists}). In both cases, their action distributions are biased to lower values relative to the mean of the distribution and display larger spreads. 
It is possible that the MLE procedure picks up on this bias and larger action spread to produce a set of basis coefficients that return a mass profile that is lower than expected.

\subsection{Energy clustering - Milky Way mass profile}\label{sec:obslike-constraints-energy}

\begin{figure*}
    \centering
    \includegraphics[width=1\linewidth]{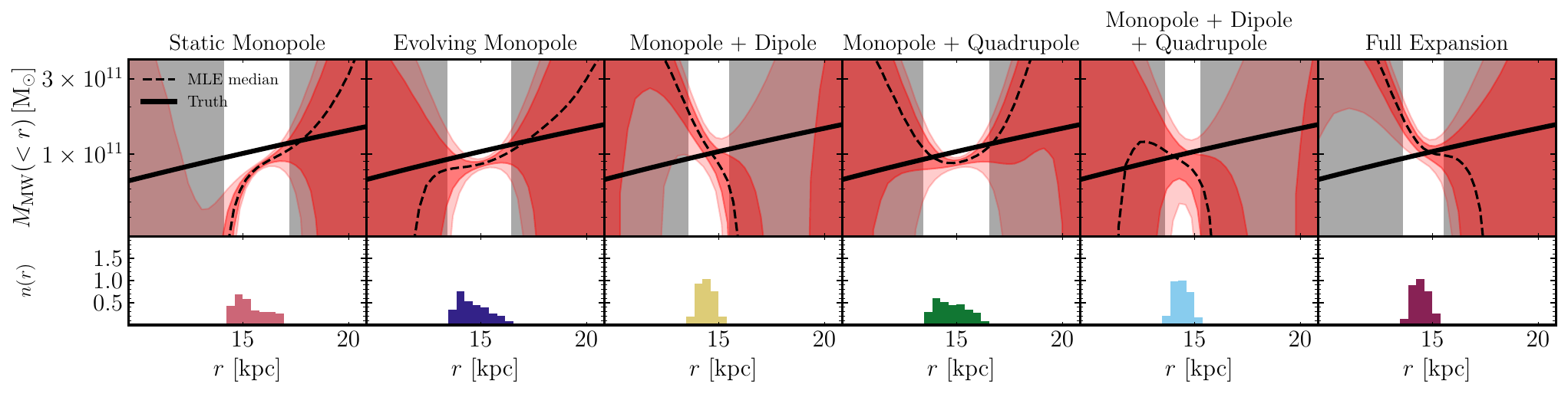}
    \includegraphics[width=1\linewidth]{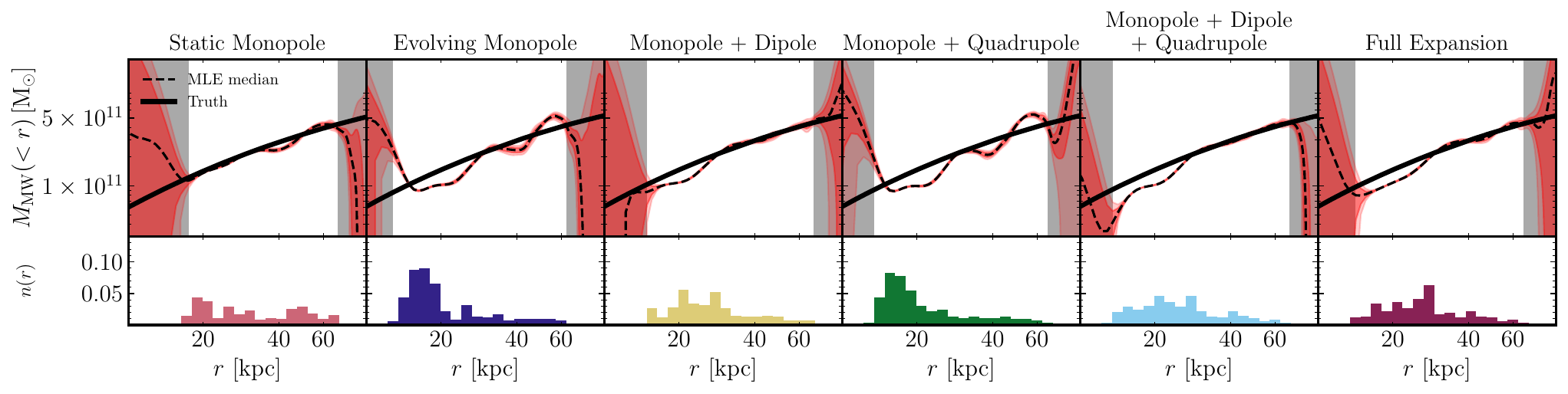}
    \caption{Constraints on the spherically-averaged Milky Way mass profile using the generated GD-1 streams (top row) and generated OC streams (bottom row) from their \textbf{energy distribution}. A multivariate normal sampling of the maximum likelihood monopole basis function expansion coefficients from the Cramér--Rao covariance matrix is carried out to obtain the median mass profile (black dashed lines) with the $1\,\sigma$ (dark red) and $2\,\sigma$ (light red) confidence intervals as shaded bands. The number density of stream members as a function of radius is shown in the lower panels, the total number of stream members matches observational counts in Sec.~\ref{sec:mockstream-generation}. The shaded-grey regions indicate radial regions where there are no stream members. The thick black line represents the true MW halo monopole potential governing the spherical part of the basis function expansion. Left to right shows these mass profiles for the streams generated in the fully live LMC + MW halo harmonic subset potentials: static monopole, evolving monopole, monopole + dipole, monopole + quadrupole, monopole + dipole + quadrupole and the full expansion. Streams that cannot recover the mass profile across their radial extent suggest their energies have been impacted by the time-dependence of the MW halo potential.}
    \label{fig:Mass-enclosed-profiles-energies}
\end{figure*}

The entire action clustering method presented so far can be replicated by replacing the actions of stream members with their energies. The energies can be simply described as the sum of the kinetic and potential energy. By assuming that the energies of stream members should be normally distributed, we can find similar sets of MLE monopole coefficients that best describe the gravitational potential given the energy distribution of the streams generated in the various time-dependent potentials. In the case of energy, each element of the FIM has a slightly different calculation and is detailed in Appendix~\ref{app:D}. 
Energy clustering is expected to be sensitive even to adiabatic changes to the potential. Hence we could expect energy clustering methods to ‘break down' faster than action clustering. Further, spherical radial actions will change in non-spherical potentials regardless of the adiabatic state of the potential. Whereas, energies do not suffer this problem but are more sensitive to time dependence.

Using these spherically averaged coefficients, we calculate the energies and quantities required to determine the FIM. Again, to make the FIM realistic in connection with observation, we make conservative matches for the counts of likely stream members in the same fashion as the actions in Sec.~\ref{sec:mockstream-generation}. Once we have the energy FIM, we take its inverse to return the energy Cramér--Rao covariance matrix. Again, we draw random samples of spherically averaged coefficients from a multivariate normal distribution with the mean being the MLE coefficients and the covariance matrix being the Cramér--Rao matrix. Using these samples of MLE coefficients in combination with the force basis function weights, we compute the MW mass profiles for the GD-1 streams (top row Fig.~\ref{fig:Mass-enclosed-profiles-energies}) and OC streams (bottom row Fig.~\ref{fig:Mass-enclosed-profiles-energies}). Fig.~\ref{fig:Mass-enclosed-profiles-energies} mimics Fig.~\ref{fig:Mass-enclosed-profiles} but for the constraints made by using the stream energies instead of radial actions. 

For the GD-1 streams, top row Fig.~\ref{fig:Mass-enclosed-profiles-energies}, we can recover the mass profile across a portion of the stream within $2\,\sigma$ in all cases. The deformations of the MW halo in its monopole and quadrupole harmonics are the mildest \citep[see fig.~5,][]{2023MNRAS.518..774L} and we could expect that the energies are the least affected by their inclusion in the potential. However, these deformations seem to affect the energies in such a way that the recovered spherically averaged mass profile is underestimated across portions of the stream radial range. 

For the OC streams, bottom row Fig.~\ref{fig:Mass-enclosed-profiles-energies}, we obtain similar results to the action clustering method with less obvious deviations when including the time-dependent MW halo harmonics. For the static monopole case, we can recover the mass profile within $2\,\sigma$ across the stream range. 
However, similarly to action clustering, we are unable to recover the mass profile across the full stream range when including time dependence in the MW halo potential. The nonphysical negative mass dips seen in Fig.~\ref{fig:Mass-enclosed-profiles-energies} are damped in comparison to the same dips seen in the mass profile from action clustering, Fig.~\ref{fig:Mass-enclosed-profiles}. This is a positive result if one wishes to measure the mass profile of galaxies using stream clustering methods when a system is in disequilibrium. 

\section{Discussion}\label{sec:discussion}

\subsection{Context of results}\label{sec:context-results}

We have demonstrated that OC-like streams generated in MW--LMC potentials including \textit{any} deformations to the MW halo will sufficiently break down action clustering, such that we cannot locally recover the spherically averaged mass profile, Fig.~\ref{fig:Mass-enclosed-profiles}. Whereas, for GD-1-like streams, we are still able to locally recover the mass profile even when the MW halo is allowed to be fully deforming. These results highlight the importance of considering deformations to the Galactic potential when modelling streams which are hotter, longer and near the LMC e.g. OC. 

The leading order deformation to the MW halo is the dipole harmonic \citep{2023MNRAS.518..774L}, i.e. the displacement in the MW halo centre due to the LMC’s gravitational effect. This could imply that a better frame of reference for evaluating the actions is the shared centre of mass frame. Re-centring the MW--LMC system could remove the non-adiabatic behaviour that is implied by the stream actions, while simultaneously offering a possible explanation as to why energy clustering seems to be less sensitive to the halo deformations. 

We showed that a similar analysis can be carried out using the clustering of the stream energies. We found tentative evidence that energy clustering is less susceptible to MW halo deformations as the deviations in the spherically averaged mass profile are damped with respect to the results from action clustering, Fig.~\ref{fig:Mass-enclosed-profiles-energies}. 
\citet{2012ApJ...760....2P} used the energies of stream members in a distinct statistical technique to constrain the MW potential by the minimization of entropy. This method is related to ours, although our Fisher information approach is more clearly related to Bayesian statistics. This work is the first formalism of using Fisher information to determine the model uncertainties when using a time-dependent BFE model of the gravitational potential while acting as a complementary effort to other studies pushing information theory into time-dependency (Lilleengen et al.~\textit{in prep}, Erkal et al.~\textit{in prep}).
Via this approach, the accuracy in recovering the MW mass profile is sensitive to where stream members exist in the Galaxy, i.e. a localised constraint \citep{2018ApJ...867..101B}, and any model assumptions made e.g. Gaussian distributions for the actions. Current state-of-the-art MW mass estimates using streams have extrapolated mass enclosed estimates further out in the MW halo to the virial radius \citep{Wang2020, 2021MNRAS.501.2279V, Reino2021, Reino2022, Ibata2024}. However, any constraint using a non-parametric description for the potential, e.g. a BFE, can only produce a localised constraint. 
Plus, for a Fisher information approach, the precision on the returned mass profile is controlled by the number of stars observed in a stream, their associated uncertainties in their positions/velocities, and the intrinsic stream width. This will vary on a stream-by-stream basis. Recent review papers for MW \citep[e.g.,][]{Wang2020, Bonaca2024} show that we know the mass to a precision of $\sim10$~per~cent where we have visible tracers.

\subsection{Caveats}\label{sec:caveats}

Our action clustering method contains sources of bias that are unaccounted for in our model. The first is biases introduced due to the energy (phase) sorting of stars along stellar streams. For individual streams, maximal clustering can occur for the wrong potential because we do not include action--phase information. Neglecting the phase information could in principle find a potential that exactly cancels action--angle correlations, producing a more tightly clustered action distribution than that for the true potential. The bias on the potential will differ for each stream and will likely cancel when considering populations of streams simultaneously \citep{2015ApJ...801...98S, Reino2021}. 

Another source of bias is due to the energy bi-modality of stars in stellar streams. A bias which also affects entropy-based techniques \citep{2012ApJ...760....2P}. As stars are stripped from the progenitor's Lagrange points during its orbit in the MW halo, they form two distinct tidal tails; the leading and trailing stream arms. If sufficiently separated, the leading and trailing tails can have distinct energy distributions (i.e., they do not overlap in energy space) with orbital energies that are higher and lower than that of the progenitor, respectively \citep{Eyre2011, 2012ApJ...760....2P}. Similarly to phase sorting, this effect translates into action space as the radial action depends on the energy of stars producing a ‘clumps within clumps' effect. 

Throughout this work, we are limited by the necessity to spherically average the BFE coefficients given our use of spherical actions. Future work to extend the current formalism to recover asymmetries in the MW--LMC system would require a larger set of basis coefficients to be constrained i.e. the harmonic orders $l > 0$. Such an approach could improve the recovered properties, but it would require using axisymmetric actions. Further extensions could be to include the conjugate angles and the MW disc in the potential.

Finally, there is possible insensitivity of the action clustering due to non-adiabatic perturbations to the potential. Given a stream that is clumped in phase space, i.e. a short and cold stream, it is possible that large-scale and low harmonic order deformations to the potential, i.e the lowest order radial functions and the dipole/quadrupole of the BFE, respectively, could be non-adiabatic but will affect the actions of all stream members in the same way. This would shift the entire distribution of stream actions without causing the clustering to disperse. Hence, non-adiabatic changes to the potential could still allow action clustering methods to work. This is most likely for the coldest and shortest streams in the Galaxy. Hotter and longer streams are likely to show dispersion in their clustering when there are non-adiabatic changes to the potential. 

\section{Conclusions}\label{sec:summary}

The merger event of the LMC with the MW is causing significant disruption in the system \citep[e.g.][]{2019ApJ...884...51G, 2019MNRAS.487.2685E, 2020MNRAS.494L..11P, 2021Natur.592..534C}, in particular, the deformations of both the MW and LMC dark matter haloes \citep{2022MNRAS.510.6201P, 2023MNRAS.518..774L}. Stellar streams in the MW will be affected \citep{2019MNRAS.487.2685E, 2019MNRAS.485.4726K, 2021ApJ...923..149S, 2023MNRAS.518..774L, 2023MNRAS.tmp..536K}. The clustering of stream actions has been used to constrain the mass profile of the MW when the potential is assumed to be static or adiabatically time-dependent \citep{2015ApJ...801...98S, Yang2020, Reino2021, Reino2022}. When time dependence is introduced into the potential in the form of galaxy mergers, the clustering of actions is subject to biases \citep{2022ApJ...939....2A}. The deformations to the MW dark matter halo due to the LMC are an example of such a system. Whether these deformations perturb the potential in an adiabatic way is unknown and would impact upon using action clustering to constrain the MW mass enclosed profile. 

We have demonstrated the ability of action clustering methods to constrain the MW mass profile by using the $N$-body simulations of \citet{2023MNRAS.518..774L} which model the deforming MW--LMC system using a BFE description using the \textsc{exp} toolkit \citep{2022MNRAS.510.6201P}. We use the spherical action clustering of GD-1 and OC streams generated in various MW--LMC potentials to infer the mass profiles. This allows us to investigate which harmonic modes of the MW halo become sufficiently non-adiabatic such that we are unable to recover the mass profile. Our uncertainties are provided using an information theory approach. This is the first time such a formalism has been used for a BFE description of the MW--LMC potential.  

Our main conclusions are:

\begin{enumerate}
    \item Using the action clustering of GD-1-like streams, i.e. cold, globular cluster streams well separated from the LMC, we can recover the mass profiles within $2\,\sigma$ no matter the level of deformations to the MW halo. This implies that any time dependence is adiabatic over their evolution, allowing action clustering of these streams to be used to infer the mass profile of the MW halo.
    \item Using the action clustering of OC-like streams, i.e. hot, dwarf galaxy streams close to the LMC, the inclusion of \textit{any} time-dependence in the MW halo potential leads to an inability to recover the mass profile within $2\,\sigma$. This suggests deformations to the MW halo introduce non-adiabatic behaviour in the potential that is sustained throughout the evolution of an OC-like stream.
    \item Using the energy clustering of GD-1-like streams, we can recover the mass profiles within $2\,\sigma$. 
    \item Using the energy clustering of OC-like streams, we find similar results to that from action clustering. Although, the deviations away from the expected mass are not as extreme. 
    \item All mass-profile constraints made using action or energy clustering are only local to the radial extent of the stream. 
\end{enumerate}

Our results have demonstrated using action clustering methods to constrain the Galaxy properties when the MW halo is deforming due to the merger with the LMC. An interesting takeaway is that the energies of the streams seem to be less strongly affected, particularly for the OC stream. Recent observational studies using stellar streams hosted around external galaxies have been able to constrain the mass distribution of the host galaxy \citep{Pearson2022a, Pearson2022b, Nibauer2023}. As the prospect of detecting more low surface brightness streams in external galaxies is set to increase with the \textit{Nancy Grace Roman Space Telescope} \citep{Spergel2015}, the increased number of streams opens up the exciting prospect for using energy clustering of external streams as a method to measure the masses of other galaxies within the Local Volume as well. To achieve this, the next step would be to apply the current methodology to phase space data with missing information, e.g., without distances to the stream.

\section*{Acknowledgements}

We thank the anonymous referee for a careful review of the manuscript that led to numerous improvements. RANB acknowledges support from the Royal Society and would like to thank Akshara Viswanathan, Ralph Schoenrich and Denis Erkal for insightful discussions. JLS acknowledges support from the Royal Society (URF\textbackslash R1\textbackslash191555). MSP is supported by a UKRI Stephen Hawking Fellowship. AP is supported by funding from the European Union’s Horizon 2020 research and innovation programme under grant agreement No. 818085 GMGalaxies. 
We used open-source software: Astropy \citep{astropy:2013, astropy:2018, astropy:2022}, NumPy \citep{harris2020array}, \textsc{jax} \citep{jax2018github}, SciPy \citep{2020SciPy-NMeth}, Matplotlib \citep{Hunter:2007}, Eigen \citep{eigenweb}, \textsc{exp} \citep{2022MNRAS.510.6201P}, \textsc{agama} \citep{2019MNRAS.482.1525V}, and galpy \citep{2015ApJS..216...29B}.

\section*{Data Availability}

A python interface to integrate orbits and access to the expansion model for the simulation can be found here: \url{https://github.com/sophialilleengen/mwlmc}.

\bibliographystyle{mnras}
\bibliography{paper} 

\appendix

\section{Milky Way dark matter halo density contrast}\label{app:A}

\begin{figure*}
    \centering
    \includegraphics[width=\linewidth]{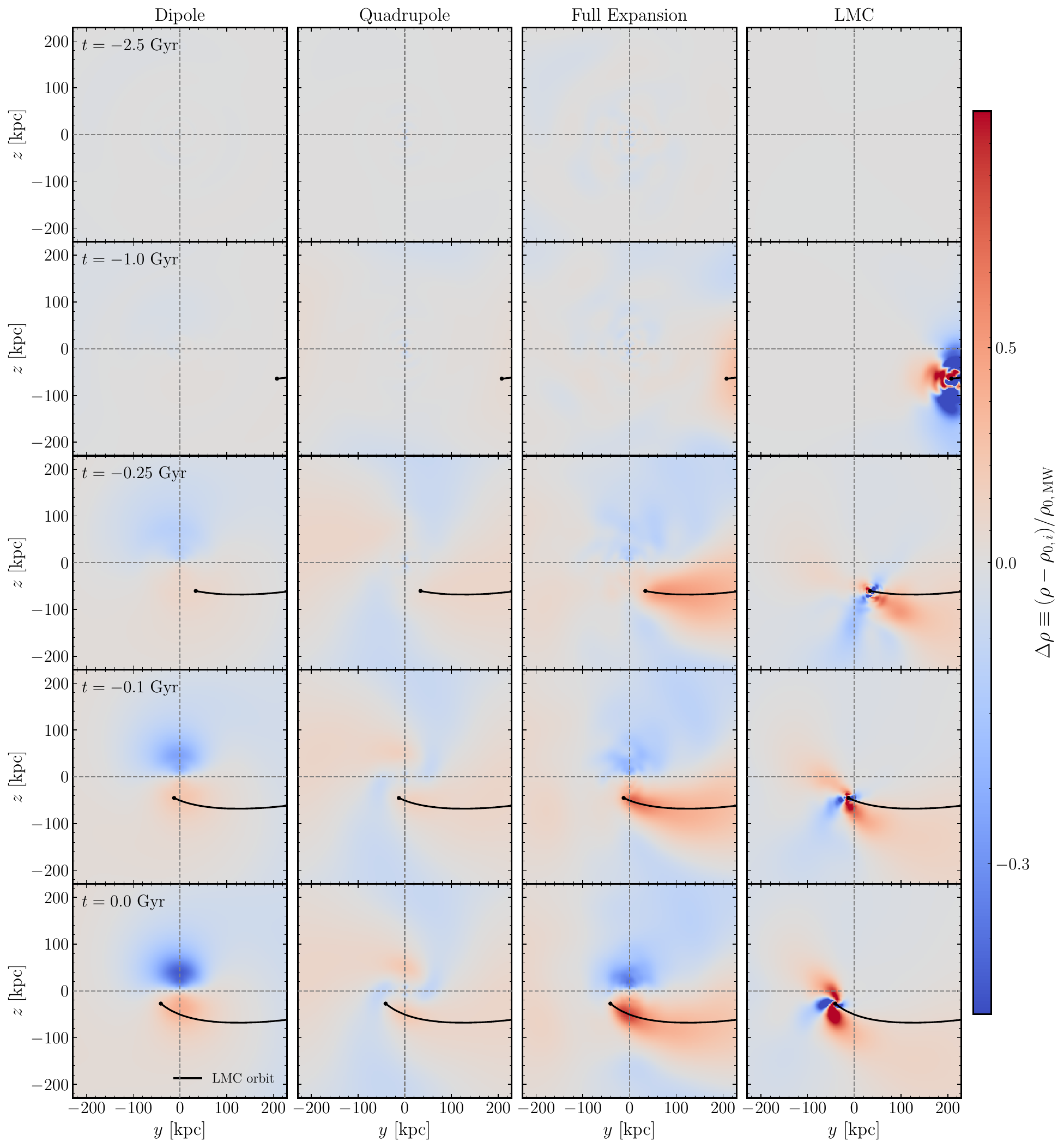}
    \caption{Temporal development of various MW dark matter halo harmonics and the LMC over the live simulation time; $t = -2.5 \,\mathrm{Gyr}$ to $t = 0 \,\mathrm{Gyr}$ with time increasing top to bottom. Going left to right across the columns shows the MW dipole, MW quadrupole, MW full expansion harmonic, and LMC full expansion. The densities are computed in the $x = 0$ Galactocentric plane in a slab of 10 kpc thickness. The colour map represents the density contrast, $\Delta\rho \equiv (\rho - \rho_{0,\,i}) / \rho_{0,\,\mathrm{MW}}$, where $\rho_{0, i}$ corresponds to the monopole density computed using only the $l=0$ order of either the MW (first three columns) or LMC expansion (final column). The track of the LMC through this plane is shown as the black line. Halo deformations due to the MW disc are omitted as they are subdominant with respect to the outer halo deformations.} 
    \label{fig:harmonics-density}
\end{figure*}

In Fig.~\ref{fig:harmonics-density}, we demonstrate the temporal development of the MW halo density contrast due to the LMC's passage for both isolated harmonic subsets and the full basis expansion simulation in the MW--LMC simulations of \citet{2023MNRAS.518..774L}. This figure replicates fig.~\ref{fig:harmonics-potential} using the densities instead of the potentials.

\citet{2019ApJ...884...51G, 2021ApJ...919..109G} present similar, yet distinct, cold dark matter simulations of the MW--LMC system and highlighted a similar scenario for the density contrast at present-day in the latter's fig.~1f. They state that the LMC imposes effects on the MW that are threefold: the collective response is primarily due to the shift of the inner halo relative to the outer halo; a global underdensity surrounds the transient response; and the transient response itself. The strength of the collective response density contrast at present-day is much higher in these simulations than in the ones considered in this work \citep{2023MNRAS.518..774L}. However, this discrepancy can be explained given the differences between the two MW--LMC simulations: first, the degree of the system's radial anisotropy will cause orbits of simulation particles to vary and re-distribute them. \citet{2021ApJ...919..109G} explored the possibility of radially biased and isotropic MW kinematics, although both have similar effects on the inner halo at radii $< 30 - 50\,\mathrm{kpc}$, i.e. their fig.~15. Secondly, the mass of the LMC in \citet{2021ApJ...919..109G} is around 5-6 times more massive than \citet{2023MNRAS.518..774L} with the former finding the strength of the $l = 1$ term (dipole) to be most impacted by varying the LMC mass. The adopted LMC mass affects the density distribution, which translates into characteristic visible changes to the stellar halo distribution \citep{2023arXiv230604837V, Foote2023}. Other subdominant differences include the resolution of the dark matter particles and basis expansion. All of the above can impact the final density distribution and strength of the LMC's dynamical friction properties. Future work devoted to understanding the extent to which the properties of the LMC, such as its mass and orbital trajectory, affect the strength of its dynamical friction signature is crucial to fully understanding the recent merger.

\section{Adiabatic invariants}\label{app:B}

This derivation is based on concepts outlined in \citet{Landau1969}, \citet{Wells2007} \& \citet{, 2008gady.book.....B}. 

Consider a system with a potential $\Phi(\mathbf{x}; \lambda(t))$. This potential is a function of the time-dependent parameter $\lambda(t)$ such that the energy is no longer conserved i.e. $E = E(t)$,

\begin{equation}\label{equ:B1}
    \dot{E} = \frac{\partial H}{\partial \lambda} \dot{\lambda},
\end{equation}

\noindent
where the dotted notation indicates a time-derivative. There are some combinations of $E$ and $\lambda$ that will remain constant. These are called \textit{adiabatic invariants}. The actions, $J$, are functions of energy $E$ and the time-dependent parameter, $\lambda$. Varying either of these will change $J$ as:

\begin{equation}\label{equ:B2}
    \dot{J} = \frac{\partial J}{\partial E} \bigg|_{\lambda} \dot{E} + \frac{\partial J}{\partial \lambda}\bigg|_{E} \dot{\lambda}
\end{equation}

\noindent
An adiabatic invariant is when $\dot{E}$ and $\dot{\lambda}$ are related in such a way that the two terms in Equ.~\ref{equ:B2} cancel. These two terms can be dealt with individually and be written as:

\begin{equation}\label{equ:B3}
    \frac{\partial J}{\partial E} \bigg|_{\lambda} = \frac{1}{\Omega} = \frac{T}{2\pi}, 
\end{equation}

\begin{equation}\label{equ:B4}
    \frac{\partial J}{\partial \lambda}\bigg|_{E} = -\frac{1}{2\pi}\int_{0}^{T} \frac{\partial H}{\partial \lambda} \bigg|_{E} dt' 
\end{equation}

\noindent
Where $\Omega$ is the frequency of an orbit in the system, and $T$ is the corresponding time period. The final result can be found by combining equs.~\ref{equ:B1}-\ref{equ:B4} to give:

\begin{equation}\label{equ:B5}
    \dot{J} = \frac{1}{\Omega}\bigg[\frac{\partial H}{\partial \lambda}\bigg|_{E} - \frac{1}{T}\int_{0}^{T} \frac{\partial H}{\partial \lambda} \bigg|_{E} dt' \bigg] \dot{\lambda}
\end{equation}

Now, we can make Equ.~\ref{equ:B5} specific to our analysis. The time-dependent parameter $\lambda(t)$ is replaced by the basis function coefficient, $A_{\mu}(t)$. This makes the partial derivative of the Hamiltonian with respect to the time-dependent parameter straightforward as we know this derivative to be, $\partial H / \partial A_{\mu} = \phi_{\mu}(\textbf{x})$ from Equ.~\ref{equ4}. This makes Equ.~\ref{equ:B5} read as:

\begin{equation}\label{equ:B6}
    \dot{J} = \frac{1}{\Omega}\bigg[\sum_{\mu}\phi_{\mu}(\mathbf{x}) - \frac{1}{T}\int_{0}^{T} \sum_{\mu}\phi_{\mu}(\mathbf{x})\, dt' \bigg] \dot{A_{\mu}}
\end{equation}



\begin{figure}
    \centering
    \includegraphics[width=\linewidth]{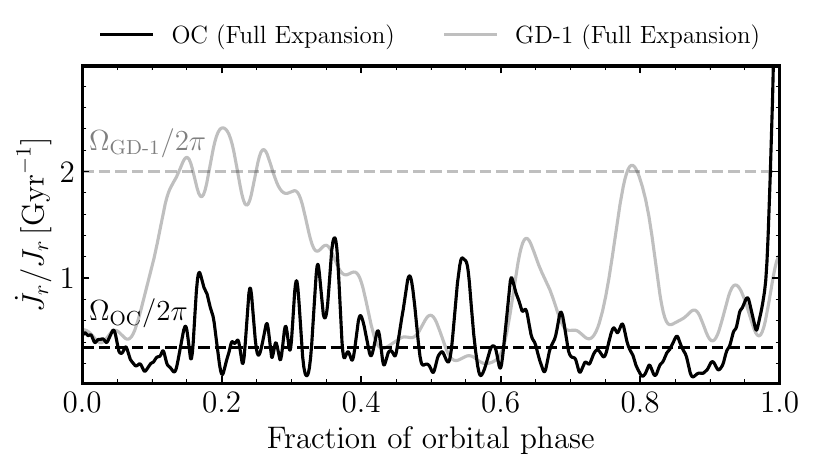}
    \caption{The instantaneous time derivative of the radial action calculated using Equ.~\ref{equ:B6}, normalised by the action of the progenitor at each time, $\dot{J_r}/J_r$, for the OC (black) and GD-1 (grey) streams in a full-expansion, time-dependent MW--LMC potential over their last full orbit. We determine the fraction of the orbit that is subject to `significant' action evolution, i.e. $\dot{J_r}/J_r \gtrsim \Omega / 2\pi$ (dashed lines, same colours). For spherical actions, significant action evolution can arise from non-spherical and/or non-adiabatic evolution of the potential. We find the fraction of the orbit that is subject to non-adiabatic/non-spherical changes in the potential for the OC and GD-1 streams as $\sim50$\% and $\sim10$\%, respectively.}
    \label{figB1}
\end{figure}

\noindent
Where $\dot{A_{\mu}}(t)$ is the time derivative of the basis function coefficients. The ratio $\dot{A_{\mu}}(t) / \Omega$ is a global indicator of adiabaticity in the potential considered. The bracketed terms modulate this global quantity to the location of the particles. The first term in the brackets is the variation in particle energy. The second term is an integral of the changes in basis function over the orbital time period of a particle. Cancellation of these two terms for the integral around an orbit gives rise to adiabatic invariance. 
In Fig.~\ref{figB1}, we evaluate Equ.~\ref{equ:B6} for an orbit of the OC and GD-1 streams evolved in the full expansion MW halo potential. We define `significant' action evolution as $\dot{J_r}/J_r \gtrsim \Omega / 2\pi$ i.e. the radial action will change by itself over an orbital period. As we are evaluating the spherical action, significant action evolution arises from non-spherical and/or non-adiabatic evolution of the potential. We find that the fraction of the orbit that is subject to significant action evolution
for the OC and GD-1 streams are $\sim50$~per~cent and $\sim10$~per~cent, respectively. This agrees with the present-day action distributions in Fig.~\ref{fig:Jr_dists} for each stream generated in the full expansion MW halo potential. The GD-1 stream has a well-clustered action distribution as expected for only adiabatic spherical changes to the potential, but the OC stream is multi-modal, hinting at non-adiabatic/non-spherical changes in the potential.

As noted, the computation of the radial action only uses the spherical terms of the BFE i.e. the monopole terms, so neglects non-spherical contributions. To evaluate the relative importance of the action evolution from non-adiabatic and non-spherical contributions, we determine Equ.~\ref{equ:B6} for the evolving monopole MW halo potential. As this potential is spherical, all action evolution can be attributed to the non-adiabatic time dependence of the potential. We find significant action changes for the OC and GD-1 streams over $\sim50$~per~cent and $\sim2.5$~per~cent of the orbits, respectively. For OC, this implies most, if not all, of the action evolution is driven by non-adiabatic changes in the potential. For GD-1, this suggests that the action evolution is mainly driven by the non-spherical evolution of the potential, while contributions from non-adiabatic changes are negligible.

\section{Derivation of Fisher Information Matrix elements - Actions}\label{app:C}

\subsection{General distribution}

Given the observation of an ensemble of particles, for $N$ model parameters so that $\mathbf{a} = [a_1, a_2, ..., a_N]^{\mathrm{T}}$ that define the action $J$, the Fisher information is given by the $\mathrm{N}\times\mathrm{N}$ positive semi-definite matrix called the \textit{Fisher Information Matrix (FIM)}: 

\begin{equation}\label{equ:C1}
\begin{split}
    [I(\mathbf{a})]_{i,j} = & \, \mathbb{E}\left[ \left( \frac{\partial}{\partial a_i} \ln f(J;\mathbf{a}) \right) \left(\frac{\partial}{\partial a_j} \ln f(J;\mathbf{a}) \right) \bigg| \,\mathbf{a}\right] \\ 
    \\
   = & \sum_{\mathrm{particles}}\left[ \left( \frac{\partial}{\partial a_i} \ln f(J;\mathbf{a}) \right) \left(\frac{\partial}{\partial a_j} \ln f(J;\mathbf{a}) \right) \right],
\end{split}
\end{equation}

\noindent
where $[I(\mathbf{a})]_{i,j}$ is the information about the model parameters $a_i$ and $a_j$ given the action $J$. The choice of the distribution of actions $f(J;\mathbf{a})$ is arbitrary. In the following section, we demonstrate its application to the Gaussian distribution.

\subsection{Gaussian distribution}

For a case of a Gaussian distribution of radial actions centred on a mean action $\langle J_{r}\rangle$ with standard deviation, $\sigma_{J_r}$ i.e., $f(J_r;\mathbf{a})= \mathcal{N}(\langle J_{r}\rangle,\,\sigma_{J_r}^{2})$, the log-likelihood is:

\begin{equation}\label{equ:C2}
    \ln{\left(f(J_r;\mathbf{a})\right)} = -\frac{1}{2}\left(\ln{(2\pi\sigma_{J_r}^2)} + \frac{\left(J_{r}(\mathbf{a}) - \langle J_{r}(\mathbf{a})\rangle\right)^{2}}{\sigma_{J_r}^2}\right)
\end{equation}

\noindent
From Equ.~\ref{equ:C1}, the Fisher information element for the $(i^{\mathrm{th}}, j^{\mathrm{th}})$ combination of parameters is:

\begin{equation}\label{equ:C3}
\begin{split}
    [I(\mathbf{a})]_{i,j} = & \sum_{\mathrm{particles}} \bigg[ \left(\frac{\partial J_{r}(\mathbf{a})}{\partial a_i} - \frac{\partial \langle J_{r}(\mathbf{a})\rangle}{\partial a_i}\right) \left(\frac{\partial J_{r}(\mathbf{a})}{\partial a_j} - \frac{\partial \langle J_{r}(\mathbf{a})\rangle}{\partial a_j}\right)
    \\
    & \times \frac{\left(J_{r}(\mathbf{a}) - \langle J_{r}(\mathbf{a})\rangle\right)^{2}}{\sigma_{J_r}^{4}} \bigg]
\end{split}
\end{equation}

\noindent
We now replace the general Gaussian distribution $f(J_r;\mathbf{a})$ with the model-specific distribution. Our radial action $J_r$ has the model variables of phase-space coordinates and is parameterised by the basis function expansion coefficients, i.e., $\mathbf{a} = [A_{0}, A_{1}, ..., A_{N}]^{T} = \{A_{\mu}\}$, such that its distribution is described by: $J_r = J_{r}(\mathbf{x}, \mathbf{v}, \{A_{\mu}\})$. Using Leibniz's rule for differentiation, the derivative of the radial action with respect to the BFE coefficients is:

\begin{equation}\label{equ:C4}
\begin{split}
    \frac{\partial J_r}{\partial A_{i}} & = \frac{1}{\pi}\int_{r_p}^{r_a} dr \frac{\phi_{i}(r_0) - \phi_{i}(r)}{\left(2E - 2\Phi(r) - L^2/r^2 \right)^{1/2}} \\
    \\
    & = \frac{\phi_{i}(r_0)}{\Omega_r} - \mathrm{I}(\textbf{J})
\end{split}
\end{equation}

\noindent
Where $\phi_{i}$ is the basis function evaluated at the particle's position $r_0$, the frequency is $\Omega_r = dH/dJ_{r}$ and I(\textbf{J}) is some integral (constant) which is the same for all particles with the same action. Now we can substitute Equ.~\ref{equ:C4} into Equ.~\ref{equ:C3} to give:

\begin{equation}\label{equ:C5}
\begin{split}
     [I(\mathbf{A_{\mu}})]_{i,j} = & \sum_{\mathrm{particles}} \bigg[ \frac{\left(\phi_{i}(r_0) - \langle\phi_{i}(r_0)\rangle\right) \left(\phi_{j}(r_0) - \langle\phi_{j}(r_0)\rangle\right)}{\Omega_{r}^{2}}
     \\
     & \times \frac{\left(J_{r} - \langle J_{r}\rangle\right)^{2}}{\sigma_{J_r}^{4}} \bigg]
\end{split}
\end{equation}

\noindent
Which is as per the expression given in Sec.~\ref{sec:fisherinfo}, Equ.~\ref{equ7}. 

\section{Derivation of Fisher Information Matrix elements - Energies}\label{app:D}

\subsection{General distribution}

Given the observation of an ensemble of particles, for $N$ model parameters so that $\mathbf{a} = [a_1, a_2, ..., a_N]^{\mathrm{T}}$ that define the energies $E$, the \textit{FIM} is: 

\begin{equation}\label{equ:D1}
\begin{split}
    [I(\mathbf{a})]_{i,j} = & \, \mathbb{E}\left[ \left( \frac{\partial}{\partial a_i} \ln f(E;\mathbf{a}) \right) \left(\frac{\partial}{\partial a_j} \ln f(E;\mathbf{a}) \right) \bigg| \,\mathbf{a}\right] \\ 
    \\
   = & \sum_{\mathrm{particles}}\left[ \left( \frac{\partial}{\partial a_i} \ln f(E;\mathbf{a}) \right) \left(\frac{\partial}{\partial a_j} \ln f(E;\mathbf{a}) \right) \right],
\end{split}
\end{equation}

\noindent
where $[I(\mathbf{a})]_{i,j}$ is the information about the model parameters $a_i$ and $a_j$ given the energy $E$. The choice of the distribution of energies $f(E;\mathbf{a})$ is arbitrary. In the following section, we demonstrate its application to the Gaussian distribution.

\subsection{Gaussian distribution}

For a case of a Gaussian distribution of energies centred on a mean energy $\langle E\rangle$ with standard deviation, $\sigma_{E}$ i.e., $f(E;\mathbf{a})= \mathcal{N}(\langle E\rangle,\,\sigma_{E}^{2})$, the log-likelihood is:

\begin{equation}\label{equ:D2}
    \ln{\left(f(E;\mathbf{a})\right)} = -\frac{1}{2}\left(\ln{(2\pi\sigma_{E}^2)} + \frac{\left(E(\mathbf{a}) - \langle E(\mathbf{a})\rangle\right)^{2}}{\sigma_{E}^2}\right)
\end{equation}

\noindent
From Equ.~\ref{equ:D1}, the Fisher information element for the $(i^{\mathrm{th}}, j^{\mathrm{th}})$ combination of parameters is:

\begin{equation}\label{equ:D3}
\begin{split}
    [I(\mathbf{a})]_{i,j} = & \sum_{\mathrm{particles}} \bigg[ \left(\frac{\partial E(\mathbf{a})}{\partial a_i} - \frac{\partial \langle E(\mathbf{a})\rangle}{\partial a_i}\right) \left(\frac{\partial E(\mathbf{a})}{\partial a_j} - \frac{\partial \langle E(\mathbf{a})\rangle}{\partial a_j}\right)
    \\
    & \times \frac{\left(E(\mathbf{a}) - \langle E(\mathbf{a})\rangle\right)^{2}}{\sigma_{E}^{4}} \bigg]
\end{split}
\end{equation}

\noindent
We now replace the general Gaussian distribution $f(E;\mathbf{a})$ with the model-specific distribution. The particle energies $E$ have the model variables of phase-space coordinates and are parametrized by the basis function expansion coefficients, i.e., $\mathbf{a} = [A_{0}, A_{1}, ..., A_{N}]^{T} = \{A_{\mu}\}$, such that $E = E(\mathbf{x}, \mathbf{v}, \{A_{\mu}\}) = |\mathbf{v}|^{2}/2 + \sum_{\mu}A_{\mu}\phi_{\mu}(\textbf{x})$. The derivative of the particle energy with respect to the BFE coefficients is $\partial E(\mathbf{x}, \mathbf{v}, \{A_{\mu}\})/\partial A_{\mu} = \phi_{\mu}(\textbf{x})$. Substituting this derivative into Equ.~\ref{equ:D3} gives the final FIM element expression:

\begin{equation}\label{equ:D4}
\begin{split}
     [I(\mathbf{A_{\mu}})]_{i,j} = & \sum_{\mathrm{particles}} \bigg[ \left(\phi_{i}(r_0) - \langle\phi_{i}(r_0)\rangle\right) \left(\phi_{j}(r_0) - \langle\phi_{j}(r_0)\rangle\right)
     \\
     & \times \frac{\left(E - \langle E\rangle\right)^{2}}{\sigma_{E}^{4}} \bigg],
\end{split}
\end{equation}

\noindent
where $\phi_{i}$ is the basis function evaluated at the particle's position $r_0$. This expression for the energy Fisher information is closely connected to the action Fisher information by the simple relations: $\Omega_r = dE/dJ_{r}$ and $\sigma_{E} = \sigma_{J_{r}}\times\Omega_{r}$

One important difference between the Fisher information expressions for the actions and energies is that for the latter, we do not have to assume that $\mathrm{I}(\textbf{J})$, Equ.~\ref{equ:D3}, is the same for all particles. This simplification makes the action spread smaller and hence we can linearly propagate the action Fisher information in Equ.~\ref{equ:C5} to get the energy Fisher information in Equ.~\ref{equ:D4}.


\bsp	
\label{lastpage}
\end{document}